\documentclass[%
pre,
twocolumn,
english,
superscriptaddress,
floatfix,
amsmath,amssymb,
longbibliography
]{revtex4-2}
\usepackage{tikz}
\usetikzlibrary{circuits.ee.IEC}
\usepackage{graphicx}
\usepackage{dcolumn}
\usepackage{soul}
\usepackage{bm}
\usepackage[hidelinks]{hyperref}
\usepackage{siunitx}
\hypersetup{colorlinks = true, allcolors = blue}

\usepackage{lipsum}

\usepackage{amssymb} 
\usepackage{amsmath} 
\usepackage{bm}
\usepackage{mathtools}
\usepackage{xcolor}

\newcommand\numberthis{\addtocounter{equation}{1}\tag{\theequation}}

\newcommand{\te}[1]{\mbox{$\mathbf{ #1 }$}}

\def\eq{\ = \ }

\def\bnabla{\boldsymbol{\nabla}}

\def\bj{\te{j}}

\def\bx{\te{x}}

\def\b1{\te{1}}

\makeatletter
\newcommand{\stepsubequation}[1][]{%
  \ifmeasuring@
  \else
    \refstepcounter{parentequation}%
    \protected@xdef\theparentequation{\arabic{parentequation}}%
    \ifdefined\theHparentequation
      \protected@xdef\theHparentequation{\arabic{parentequation}}%
    \fi
    \setcounter{equation}{0}%
    \if\relax\detokenize{#1}\relax\else
      \edef\@currentlabel{\theparentequation}%
      \ltx@label{#1}%
    \fi
  \fi
}
\makeatother

\begin{document}

\allowdisplaybreaks

\righthyphenmin=4
\lefthyphenmin=4

\title{Electrochemical response of biological membranes to \\ 
localized currents and external electric fields}

\author{Joshua B. Fernandes}
\thanks{Equal contribution}
\affiliation{Department of Chemical and Biomolecular Engineering, University of California, Berkeley, CA 94720, USA\looseness=-1}
\affiliation{Chemical Sciences Division, Lawrence Berkeley National Laboratory, CA 94720, USA}

\author{Hyeongjoo Row}
\thanks{Equal contribution}
\affiliation{Department of Chemical and Biomolecular Engineering, University of California, Berkeley, CA 94720, USA\looseness=-1}
\affiliation{Helen Wills Neuroscience Institute, California Institute for Quantitative Biosciences, QB3, Center for Computational Biology, University of California, Berkeley, CA 94720, USA}

\author{Kranthi K. Mandadapu}
\thanks{Correspondence: kranthi@berkeley.edu, kshekhar@berkeley.edu}
\affiliation{Department of Chemical and Biomolecular Engineering, University of California, Berkeley, CA 94720, USA\looseness=-1}
\affiliation{Chemical Sciences Division, Lawrence Berkeley National Laboratory, CA 94720, USA}

\author{Karthik Shekhar}
\thanks{Correspondence: kranthi@berkeley.edu, kshekhar@berkeley.edu}
\affiliation{Department of Chemical and Biomolecular Engineering, University of California, Berkeley, CA 94720, USA\looseness=-1}
\affiliation{Helen Wills Neuroscience Institute, California Institute for Quantitative Biosciences, QB3, Center for Computational Biology, University of California, Berkeley, CA 94720, USA}
\affiliation{Biological Systems Division, Lawrence Berkeley National Laboratory, Berkeley, CA 94720, USA}

\date{\today}

\begin{abstract}
Electrochemical phenomena in biology often unfold in confined geometries where micrometer- to millimeter-scale domains coexist with nanometer-scale interfacial diffuse charge layers. We analyze a model lipid membrane-electrolyte system where an ion channel-like current flows across the membrane while parallel electrodes simultaneously apply a step voltage, emulating an extrinsic electric field. Matched asymptotic expansions of the Poisson-Nernst-Planck equations show that, under physiological conditions, the diffuse charge layers rapidly reach a quasi-steady state, and the bulk electrolyte remains electroneutral. As a result, all free charge is confined to the nanometer-scale screening layers at the membrane and electrode interfaces. The bulk electric potential satisfies Laplace’s equation, and is dynamically coupled to the interfacial layers through time-dependent boundary conditions.  This multiscale coupling partitions the space-time response into distinct regimes. At sufficiently long times, we show that the system can be represented by an equivalent circuit analogous to those used in classical cable theory. We derive closed-form expressions of the transmembrane potential within each regime, and verify them against nonlinear numerical simulations.  Our results show how electrode-induced screening and confinement effects influence the electrochemical response over multiple length and time scales in biological systems.     

\end{abstract}

\maketitle

\section{Introduction}

 A defining feature of bioelectricity is that nanoscale charge separations can shape macroscopic voltage dynamics in cells and tissues~\cite{lodish2000molecular}. Cells and organelles are bounded by lipid membranes, which separate electrolytes of different compositions. Ions cross these membranes through specialized proteins such as ion channels and pumps, which transiently couple the two electrolytes and thereby regulate the transmembrane potential $V^{\rm M}$~\cite{hille1992}. Whenever an ion channel or pump moves charge across a cellular membrane, it disrupts local electroneutrality, producing an electric field. In response, diffuse charges reorganize, and their dynamics  shape both the spatiotemporal profile of $V^{\rm M}$ and the surrounding electrochemical gradients~\cite{row2025spatiotemporal}. 

Electrophysiological experiments often superpose an \emph{external} electric field---most commonly through either uniform field stimulation~\cite{bikson2004effects, radman2009role, francis2003sensitivity} or clamping protocols~\cite{neher1992patch, moore1963voltage_chapter}---by placing electrodes at fixed distances from the membrane. This electrode-membrane separation introduces additional geometry-based length and time scales that influence diffuse charge dynamics throughout the system. Unraveling how the combined influence of the imposed field and localized ionic currents shape the macroscopic electrochemical response is a problem of interest in membrane biophysics.

\begin{figure}[t!]
\centering  
\includegraphics[width=0.85\linewidth]{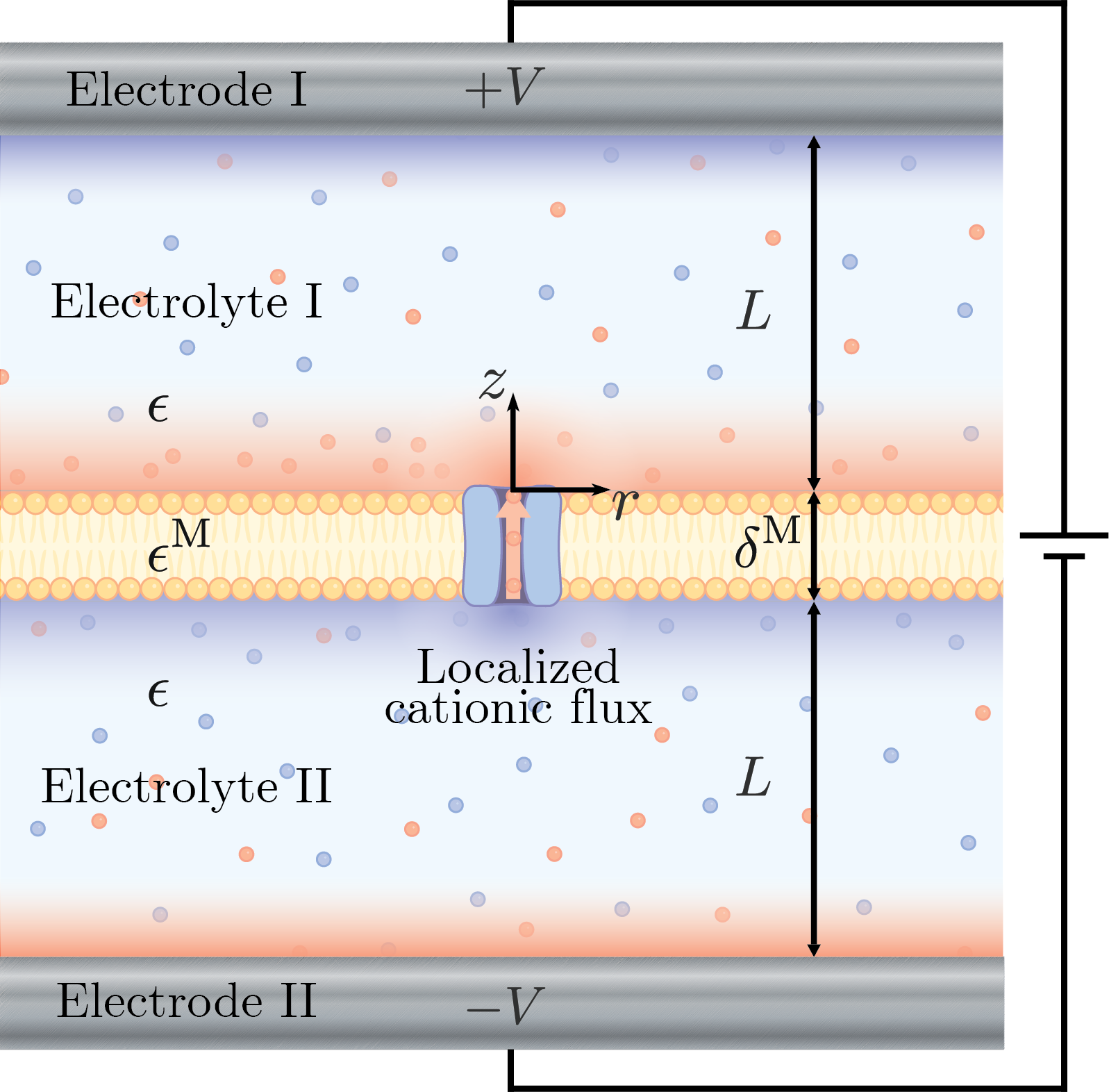}    
\caption{\textbf{Problem description.} A flat lipid membrane of thickness $\delta^{\rm M}$ lies midway between two parallel electrodes, with each gap (thickness $L$) filled with a dilute, monovalent electrolyte. Electrolyte and membrane permittivities are ${\epsilon}$ and  ${\epsilon^{\rm M}}$, respectively. The two solutions are initially homogeneous with salt concentration $C_{0}$. At ${t=0}$, the electrodes are stepped to a potential difference of $2V$, and a membrane transporter begins pumping cations selectively across the lipid bilayer. Diffuse charge layers rapidly form near the membrane and electrode surfaces (illustrated by shaded regions).  Typical physiological values for all parameters appear in Table~\ref{tab:parameters} at the end of Sec.~\ref{sec:length_time_scales}.
}
\label{fig1_schematic_descriptions}
\end{figure}

The present work builds on two recent theoretical studies from our groups, both formulated using the Poisson-Nernst-Planck (PNP) framework appropriate for dilute physiological electrolytes. In the first study, we investigated a flat biological membrane containing a single ion pump surrounded by \emph{unbounded} electrolytes on either side~\cite{row2025spatiotemporal}. As the pump fluxes ions of a single type across the membrane, we showed that the transmembrane potential $V^{\rm M}$ exhibits distinct scaling regimes as a function of radial distance $r$ from the pump. Near the pump, $V^{\rm M}$ has a monopolar profile decaying as $1/r$ until a propagating shoulder ${r^{*}=vt}$ is reached. Beyond the shoulder, $V^{\rm M}$  transitions to a dipolar $1/r^{3}$ profile. In physiological settings, the shoulder-speed ${v\sim 40 \text{ m/s}}$, so that ionic reorganization proceeds significantly faster than bare diffusive dynamics. Ultimately, such power law profiles are a telltale sign that long-ranged electrostatic fields are at play near the membrane surface. The breaking of local electroneutrality, caused by the ion-selective nature of channel of pump currents, gives rise to the long-range character.

The second study~\cite{farhadi2025capacitive} analyzed a complementary configuration consisting of an \emph{impermeable} membrane-electrolyte system sandwiched between parallel blocking electrodes. When the electrodes apply a step voltage, the system relaxes to a new equilibrium that is driven by diffuse charge reorganization in the electrolytes. At leading order, the relaxation unfolds on a \emph{capacitive timescale} that, under typical physiological settings, is roughly two orders of magnitude smaller than the bare electrolyte $RC$ timescale reported in previous works~\cite{macdonald1974binary,kornyshev1981conductivity, Bazant04, janssen2018transient}. The faster relaxation is due to the lipid membrane: its finite thickness and small dielectric permittivity relative to water reduces the system's net capacitance, which in turn lowers the $RC$ timescale, accelerating the charging and discharging.    

Here, we consider the \emph{combined} scenario sketched in Fig.~\ref{fig1_schematic_descriptions}: a planar uncharged membrane of thickness $\delta^{\rm M}$ separates two symmetric monovalent electrolytes and is flanked by electrodes. The electrodes are placed at a distance $L$ away from the membrane surfaces.  At ${t=0}$, an ion pump located at the center begins transporting cations across the membrane while the electrodes impose a step voltage $2V$. Electric fields arise instantaneously throughout the system, and drive the rapid formation of diffuse charge screening layers---of characteristic thickness $\lambda_{\rm D}$ (${\sim \text{1 nm}}$), the Debye length---at the membrane and electrode interfaces. The characteristic charging time for the diffuse layers is the Debye time ${\tau_{\rm D} \equiv \lambda_{\rm D}^{2}/D}$ (${\sim \text{1 ns}}$), where $D$ is the ionic diffusivity. 

Our goal is to understand how the transmembrane potential $V^{\rm M}(r,t)$ evolves under this dual stimulus of current and voltage. At leading order, the system’s electrochemical response can be decomposed into two independent subproblems: the voltage-driven response without transmembrane current (analyzed in Ref.~\cite{farhadi2025capacitive}), and the current-driven response when the electrodes are each held at $V=0$, which is the focus of our analysis.  
We find that for early times ${t<\tau_{\rm D}}$, the spatiotemporal dynamics are well captured by a point charge approximation~\cite{row2025spatiotemporal}, under the assumption that diffuse charge layers have not yet formed. For larger times, we develop a boundary layer theory, and analyze it using the method of matched asymptotic expansions. The theory indicates that beyond ${\tau_{\rm D}}$, the bulk electrolyte remains electroneutral, with free charge confined to the diffuse charge layers. The bulk electric potential satisfies Laplace’s equation, albeit with time-dependent boundary conditions that govern how the interfacial layers charge and discharge. 

Under the boundary layer theory, the leading-order evolution of $V^{\rm M}(r,t)$ exhibits distinct spatiotemporal regimes. For ${r<L}$, the electrodes are effectively invisible over a protracted period, and $V^{\rm M}$ follows the unbounded monopole-dipole solution presented in Ref.~\cite{row2025spatiotemporal}. However, for ${r>L}$, the dynamics are significantly influenced by electrode-induced screening effects, giving rise to three regimes: (i) an \emph{electrostatic} regime (${t<\tau_{\rm D}}$), where ${V^{\rm M} \sim \exp\left(-\pi r/L \right)}$; (ii) a \emph{capacitive} regime in which the exponentially decaying $V^{\rm M}$ steadily grows in amplitude while the diffuse layers charge over the capacitive timescale introduced in Ref.~\cite{farhadi2025capacitive}; and (iii) a \emph{diffusive} regime that begins around the bare electrolyte $RC$ timescale. In this regime, the axial electric field vanishes and $V^{\rm M}$ evolves radially in a diffusive manner. This evolution is governed by an effective diffusivity ${D_{\rm eff} \sim D L /\lambda_{\rm D}}$, and follows a dynamical equation analogous to classical cable theory~\cite{dayan2001}. The long-time behavior depends on the electrode properties. For blocking electrodes, the system does not reach a steady state---continuous transmembrane current leads to indefinite charge accumulation. In contrast, for ideal Faradaic electrodes, all transported charge is consumed at the electrodes, and the system reaches a steady state. Fully nonlinear finite-element simulations confirm all analytical results. 

This paper is organized as follows. Section~\ref{sec:Model} introduces the governing equations and boundary conditions, the relevant spatiotemporal scales, and the decomposition of Fig.~\ref{fig1_schematic_descriptions} into voltage- and current-driven subproblems. Section~\ref{sec:BL_Theory} develops the boundary layer theory featuring a charge-free bulk coupled to dynamically evolving interfacial diffuse charge layers. Section~\ref{sec:Results} presents the spatiotemporal evolution of $V^{\rm M}$ in the electrostatic, capacitive, and diffusive regimes. For each regime, we provide closed-form expressions for $V^{\rm M}$ and compare it with simulations, highlighting how the electrode-membrane spacing $L$ and the system's intrinsic timescales organize the response. We compare the dynamics arising from blocking versus Faradaic electrodes and identify when a coarse-grained circuit model becomes valid. Taken together, this work provides quantitative insights into how confinement effects, interfacial ionic reorganization, and electrode properties shape the electrochemical response of biological membranes to spatially localized currents.

\section{The Mathematical Model}\label{sec:Model}
We seek to determine the electrochemical response of the system shown in Fig.~\ref{fig1_schematic_descriptions}. The ion pump is modeled as a circular patch of radius $R^{\rm P}$, and the system is axisymmetric about the pump axis with $r$ representing the radial coordinate and $z$ representing the axial coordinate. Initially, the electrolytes are homogeneous with ionic concentration ${C^{0}}$. At time ${t=0}$, an ion-selective current is imposed through the membrane patch,  while a constant external voltage $2V$ is simultaneously applied between the electrodes. Under these conditions, we wish to determine the spatiotemporal evolution of the electric potential $\phi(\bx,t)$ and the ionic concentrations $C_{+} (\bx, t)$ and $C_{-} (\bx, t)$ of cations and anions, where $\bx=(r,z)$.  

\subsection{Electrochemical Transport Equations}\label{subsec:Problem_setup}
In the dilute limit of electrolyte transport theory \cite{fong2020transport}, the spatiotemporal dynamics in each electrolyte domain is governed by the PNP equations~\cite{Nernst1888,Nernst1889,Planck1890,Bazant04}
\begin{subequations}
\label{eq:nonlinear_eqs}
\begin{align}
    \frac{\partial C_{\pm}}{\partial t}
    &\eq
    - \bnabla \cdot \bj_{\pm}
    \ , \label{eq:dim_eq_C}
    \\
    \bj_{\pm}
    &\eq
    - D \left(
        \bnabla C_{\pm}
        \pm \frac{{\rm e}}           {k_{\mathrm{B}}T}
          C_{\pm} \bnabla \phi
    \right)
    \ , \label{eq:dim_eq_ionFlux}
    \\
    -\epsilon \nabla^{2} \phi
    &\eq
    \rho
    \eq
    {\rm e} (C_{+} - C_{-} ) \ , \label{eq:dim_eq_poisson} 
\end{align}
\end{subequations}
where $D$ is the ionic diffusivity (assumed equal for all ionic species), $k_{\mathrm{B}}T$ is the thermal energy scale, ${\rm e}$ is the fundamental charge, and $\epsilon$ is the electrolyte permittivity.  We use the subscripts $+$ and $-$ to denote cations and anions, respectively. 
Equation~\eqref{eq:dim_eq_C} is the mass balance of ions, Eq.~\eqref{eq:dim_eq_ionFlux} describes the ionic flux $\bj_{\pm}$ with the first term on the right representing diffusion and the second electromigration, and Eq.~\eqref{eq:dim_eq_poisson} is Poisson's equation relating the electric potential to the charge density $\rho$ in the electrolyte.
Ion concentrations inside the membrane are negligible due to its low  permittivity~\cite{volkov2008energetics,nymeyer2008method}, and the electrostatics in the membrane is governed by
\begin{equation}
    -\epsilon^{\rm M} \nabla^{2}  \phi^{\mathrm{M}}
    \eq
    0 \ ,
    \label{eq:dim_eq_poissonM}
\end{equation}
where $\phi^{\rm M}(\bx,t)$ is the electric potential inside the membrane, and $\epsilon^{\rm M}$ is the membrane permittivity.
This introduces the dielectric mismatch between the membrane and surrounding electrolyte ${\Gamma \equiv \epsilon/\epsilon^{\rm M} \sim 20\gg 1}$.
 
\subsection{Boundary Conditions at the Electrode and Membrane Interfaces}\label{subsec:Boundary_Conditions}
The two electrodes are stepped to fixed potentials ${\phi = \pm V}$, as shown in Fig.~\ref{fig1_schematic_descriptions}. We begin by assuming that the electrodes block Faradaic currents, resulting in no-flux conditions ${\bj_{\pm} \cdot \te{e}_{z} = 0}$ at the electrode surface, where \(\mathbf{e}_z\) denotes the unit vector in the \(z\)-direction. Blocking of current leads to the accumulation of diffuse charge over the Debye length ${\lambda_{\mathrm{D}} \equiv \sqrt{\epsilon k_{\rm B}T/(2 C^{0} \mathrm{e}^2}) }$ \cite{Gouy1910, chapman1913li} at the electrode interfaces. The alternative case of non-blocking but ideal ``Faradaic'' electrodes is treated in Sec.~\ref{sec:Faradaic_main_manuscript}.

Our membrane is impermeable to ions everywhere except at the pump patch (Fig.~\ref{fig1_schematic_descriptions}). Without loss of generality, we consider a selective cation pump.
Since the characteristic pore size is ${R^{\rm P}\sim 0.5 \text{ nm} \lesssim \lambda_{\rm D}}$~\cite{moldenhauer2016effective}, the transmembrane current can be treated as a point source, i.e.,
${\mathbf{j}_+ \cdot \mathbf{e}_z = (I/{\rm e}) \delta(r)/2\pi r}$, where $I$ (constant) is the imposed transmembrane cation current. 
Anions do not cross at any location, hence ${\mathbf{j}_{-} \cdot \mathbf{e}_z = 0}$. Since our model is \emph{kinematic} in nature, detailed internal mechanisms of ion channel or pump transport are not required. 

For a membrane without immobilized surface charges, electrostatics requires the continuity of both the electric potential and the normal component of the electric displacement field \cite{Kovetz2000}
\begin{equation}
    \phi \eq \phi^{\mathrm{M}}
    \ , \quad 
    \left(
        \epsilon^{\rm M} \nabla \phi^{\rm M}
        -\epsilon \nabla \phi
    \right) \cdot \te{e}_{z} 
    \eq
    0
    \ ,
    \label{eq:membrane_elec_BC}
\end{equation}
at either membrane-electrolyte interface. Formally, Eq.~\eqref{eq:membrane_elec_BC} couples the diffuse charge dynamics in the electrolyte (Eqs.~\eqref{eq:nonlinear_eqs}) with the electrostatics inside the membrane (Eq.~\eqref{eq:dim_eq_poissonM}). Yet, at radial distances much greater than the membrane thickness (${r\gg \delta^{\rm M}}$) a scaling argument shows that axial variations of $\phi^{\rm M}$ will dominate over radial variations, and Eq.~\eqref{eq:dim_eq_poissonM} can be approximated as ${\partial^{2} \phi^{\rm M} / \partial z^{2} \approx 0}$ (see Sec.~I.3 in SM). A straightforward analysis then replaces the boundary conditions~\eqref{eq:membrane_elec_BC} with 
\begin{equation}
    \phi
    \eq
    \pm \frac{\Gamma \delta^{\rm M}}{2} \frac{\partial \phi}{\partial z}
    + \phi^{\rm M}_{\rm mid}
    \ ,
    \label{eq:membrane_elec_RobinBC} 
\end{equation}
where the positive and negative signs correspond to the upper and lower membrane surfaces in Fig.~\ref{fig1_schematic_descriptions}. Here, $\phi^{\rm M}_{\rm mid}$ is the electric potential at the membrane midplane.

With the mathematical problem stated, we now devise a boundary layer theory amenable to analytical solutions, starting with a description of the key length- and time-scales involved.

\subsection{Length and Time Scales Governing the Electrochemical Response}
\label{sec:length_time_scales}
The problem as formulated involves two intrinsic length scales, the Debye screening length ${\lambda_{\rm D} \sim 1\text{ nm}}$ and the membrane thickness ${\delta^{\rm M}\sim5 \text{ nm}}$, and a geometry-based length scale ${L \sim 1\text{ \textmu m} - 1 \text{ mm}}$ corresponding to the electrode-membrane distance. Notably, ${L}$ is significantly larger than the other two length scales, i.e., ${L/\lambda_{\rm D},\ L/\delta^{\rm M}\gg1}$. As we show below, the presence of electrodes divides in-plane features of the electrochemical response into two spatial regimes: ${r<L}$ and ${r>L}$.

Each length scale introduces a characteristic diffusion time. Diffusion over the interfacial layer occurs over the microscopic Debye time ${\tau_{\rm D}\equiv \lambda_{\rm D}^{2} / D}$ (${\sim1 \text{ ns}}$)~\cite{haynes2012crc}, while  diffusion across the bulk electrolyte occurs over ${\tau_{\rm L}\equiv L^{2} / D}$ (${\sim 1 \text{ ms}-  10^3 \text{ s}}$). At first glance it is unclear which of these two extremes governs the charging dynamics of the membrane electrolyte system in Fig.~\ref{fig1_schematic_descriptions}.
Rather, it turns out that the leading-order charging timescale of the membrane-electrolyte system is dictated by its $RC$ circuit-like behavior \cite{farhadi2025capacitive}. The electrolyte domain acts as a resistor of length $2L$ and ionic conductivity ${g\equiv D \epsilon / \lambda_{\rm D}^2}$, giving an equivalent resistance-area ${R_{\rm eq} \equiv 2L/g}$. As capacitive elements, the membrane contributes an equivalent capacitance per unit area of ${C_{\rm M} \equiv \epsilon^{\rm M}/\delta^{\rm M}}$, while each diffuse charge layer contributes ${C_{\rm D} \equiv \epsilon/\lambda_{\rm D} }$. The series capacitance per area is therefore ${ C_{\rm eq} \equiv \left[{C_{\rm M}}^{-1} + n_{\rm D} {C_{\rm D}}^{-1}\right]^{-1} }$, where $n_{\rm D}$ is the total number of diffuse charge layers in the system. This yields the charging timescale
\begin{equation}
    \tau_{\rm C}
    \ \sim \ 
    R_{\rm eq} C_{\rm eq}
    \eq
    \frac{\tau_{\rm B}}{n_{\rm D}/2 + \chi} \ ,
    \label{eq:macro_timescale}
\end{equation}
where ${\tau_{\rm B}\equiv \lambda_{\rm D}L/D }$ is the $RC$ timescale of the bare electrolyte~\cite{Bazant04} and  ${\chi \equiv \left[(\epsilon / \lambda_{\rm D}) / 2 \right] / (\epsilon^{\rm M} / \delta^{\rm M}) = \Gamma \delta^{\rm M} / (2 \lambda_{\rm D})}$ is the ratio of the diffuse layer to membrane capacitance. For a system with blocking electrodes ${n_{\rm D} = 4}$, due to the presence of diffuse layers at the membrane and electrode surfaces, while for ideal Faradaic electrodes, ${n_{\rm D}=2}$ as these electrodes do not support diffuse charge. Equation~\eqref{eq:macro_timescale} suggests that the insulating membrane shortens the charging time by a factor ${n_{\rm D}/2 + \chi}$. The reduction can be significant under typical circumstances. For example,
with ${L\sim1 \text{  mm}}$, we obtain ${\tau_{\rm B} \sim 1\text{ ms}}$ for a bare electrolyte, whereas the membrane with blocking electrodes reduces this to ${\tau_{\rm C} \sim 20 \text{ \textmu s}}$. Table~\ref{tab:parameters} summarizes the physical parameters and their typical physiological values used throughout this work.
\begin{table}[h]
\caption{\label{tab:parameters}%
The symbols used for the parameters in the article and physiological ranges of their values. ${\epsilon_0=8.85\times 10^{-12} \text{ F}\cdot\text{ m}^{-1}}$ is the vacuum permittivity. The membrane permittivity $\epsilon^{\rm M}/\epsilon_0$ typically ranges between 2 and 5 and we choose 4 as a representative value in simulations.
}
\begin{ruledtabular}
\begin{tabular}{lcc}
    Parameter & Symbol & Typical Range \\
    \hline
    Salt concentration & $C^0$ & $100-200\text{ mM}$ \cite{milo2015cell}\\
   Pore size & $R^{\rm P}$ & $~0.5 \text{ nm}$ \cite{moldenhauer2016effective} \\
    Debye length & $\lambda_{\rm D}$ & $0.5-3\text{ nm}$\\
    Membrane thickness & $\delta^{\rm M}$ & $3-5\text{ nm}$ \cite{milo2015cell}\\
    Electrode spacing & $L$ & $1\text{ \textmu m}-1\text{ mm}$ \\
    Applied voltage & $V$ & $0-2.5\text{ V}$ \cite{hodgkin1952quantitative,hille1992,tsong1991electroporation,kotnik2019membrane} \\
    Pump current & $I$ & $10^{-3}-10 \text{ pA}$ \cite{milo2015cell}\\
    Electrolyte permittivity & $\epsilon$ & $80\epsilon_0$ \cite{haynes2012crc} \\
    Membrane permittivity & $\epsilon^{\rm M}$ & $2\epsilon_0-5\epsilon_0$ \cite{huang1977theoretical,nymeyer2008method} \\
    Ion diffusivity & $D$ & $0.5-2\text{ nm}^2/\text{ns}$ \cite{haynes2012crc} \\
    Capacitance ratio & $\chi$ & $\sim 40$\\
    Debye time & $\tau_{\rm D}$ & $\sim 1\ \text{ns}$\\
    Capacitive timescale & $\tau_{\rm C}$ & $\sim 20\text{ ns}-20\text{ \textmu s}$\\
    Bare $RC$ timescale & $\tau_{\rm B}$ & $\sim 1 \text{ \textmu s}-1 \text{ ms}$\\
\end{tabular}
\end{ruledtabular}
\end{table}

In what follows, we show that following an initial electrostatic regime of time $\tau_{\rm D}$, the leading-order charging dynamics of our system is governed by both $\tau_{\rm C}$ and $\tau_{\rm B}$. The presence of these widely separated scales gives rise to the multi-regime behavior analyzed below, and also allows us to derive regime-specific asymptotic solutions.

\subsection{Linear dynamics}\label{subsec:linearization}

Because the capacitance ratio $\chi$ is large, the membrane dominates the overall capacitive response, leading to ${C_{\rm eq} \approx C_{\rm M} \sim 1 \text{ \textmu F}/\text{cm}^2}$. This relatively small total capacitance not only shortens the charging time in Eq.~\eqref{eq:macro_timescale} but also reduces the amount of charge that must cross the membrane to establish physiologically relevant transmembrane voltages (${\sim 10-100 \text{ mV}}$). The magnitude of this charge (per unit membrane area) can be conveniently estimated using the capacitor $\mbox{charge-voltage}$ relation ${Q=C V}$. For example, a 25 mV transmembrane potential can be achieved by an areal charge density of $0.0016 \text{ e}/\text{nm}^2$. Assuming that the area of a single lipid headgroup is ${\sim 0.5 \text{ nm}^2}$, this corresponds to only one excess ion per 1,250 lipid molecules. Even a physiologically large transmembrane potential of 1 V requires just one excess ion per 30 lipids. 

Consequently, bulk ionic concentrations deviate from their initial value $C^{\rm 0}$ by only a few percent. Under these conditions, the linearized Poisson-Nernst-Planck (PNP) equations accurately capture the dynamics of the system~\cite{farhadi2025capacitive}. The linear approximation also effectively describes ionic disturbances produced by a single transporter operating at its maximal current~\cite{row2025spatiotemporal}. We therefore adopt the linearized PNP equations as a foundation for our analysis and confirm its accuracy by comparison with fully nonlinear finite‑element simulations of Eqs.~\eqref{eq:nonlinear_eqs}-\eqref{eq:dim_eq_poissonM}.
In the original PNP equations, nonlinearlity arises from the electromigration term in the ionic flux (Eq.~\eqref{eq:dim_eq_ionFlux}). Linearizing about the initial concentration modifies the ionic flux to
\begin{equation}
\bj_{\pm}
    \ \approx \ 
    - D
        \bnabla C_{\pm}
        \mp \frac{g}           {2{\rm e}} \bnabla \phi
    \ , \label{eq:dim_eq_ionFlux_lin}
\end{equation}
where $g$ is the electrolyte conductivity.

The simplification in Eq.~\eqref{eq:dim_eq_ionFlux_lin} has three important consequences. First, linearization decouples the dynamics of the overall salt concentration ${C \equiv C_{+} + C_{-}}$ from those of the charge density $\rho$. The governing equations of $\rho$ and $C$, also known as the Debye-Falkenhagen equations~\cite{debyefalkenhagen}, indicate that the salt concentration follows the usual diffusion equation, ${\partial C / \partial t - D \nabla^{2} C = 0}$, being independent of $\rho$ and $\phi$. The charge density follows ${ {\partial \rho}/{\partial t} = - \bnabla \cdot \bj_{\rho} }$ with the charge flux ${\bj_{\rho} \equiv {\rm e} \left(\bj_{+} - \bj_{-}\right) = -D \bnabla \rho -g \bnabla \phi}$. With Poisson's equation~\eqref{eq:dim_eq_poisson}, the linearized charge dynamics are described by
\begin{subequations}
\begin{align}
    \frac{\partial \rho}{\partial t} - D \nabla^{2} \rho + \frac{1}{\tau_{\rm D}} \rho
    &\eq
    0
    \ ,
    \label{eq:dim_charge_bal_lin}
    \\
    -\epsilon {\nabla}^{2} \phi 
    &\eq
    \rho
    \ .
    \label{eq:dim_eq_poisson_lin} 
\end{align}
\end{subequations}
Balancing the electrostatic relaxation term ${\rho/\tau_{\rm D}}$ in Eq.~\eqref{eq:dim_charge_bal_lin} against the accumulation and diffusion terms indicates that any free charge, persists only over length scale ${\mathcal{O}(\lambda_{\rm D})}$ for time scale ${\mathcal{O}(\tau_{\rm D})}$. The initial condition is ${\rho=0}$ at ${t=0}$ and the boundary conditions are ${\bj_{\rho} \cdot \te{e}_{z} = I\delta(r)/2\pi r}$
at the membrane surface ${z=0}$ (reflecting a source of charge), and ${\bj_{\rho} \cdot \te{e}_{z} = 0}$ at the electrode surface ${z=L}$. The membrane electric potential  $\phi^{\rm M}$ is still governed by Eq.~\eqref{eq:dim_eq_poissonM}, and the boundary conditions are as in Sec.~\ref{subsec:Boundary_Conditions}.

Second, as discussed in Refs.~\cite{row2025spatiotemporal,farhadi2025capacitive}, the linearized governing equations in $\rho$, $\phi$, and $\phi^{\rm M}$ and their associated boundary conditions are odd-symmetric about the membrane midplane. Therefore, the potential at the midplane ${\phi^{\rm M}_{\rm mid}=0}$, and
Eq.~\eqref{eq:membrane_elec_RobinBC} gives ${\phi = \left( \Gamma \delta^{\rm M} /2\right) \left( \partial \phi / \partial z \right)}$ at ${z=0}$. This simplification decouples the electrolyte domain from the membrane.

\begin{figure}[t]
\centering
\includegraphics[width=\linewidth]{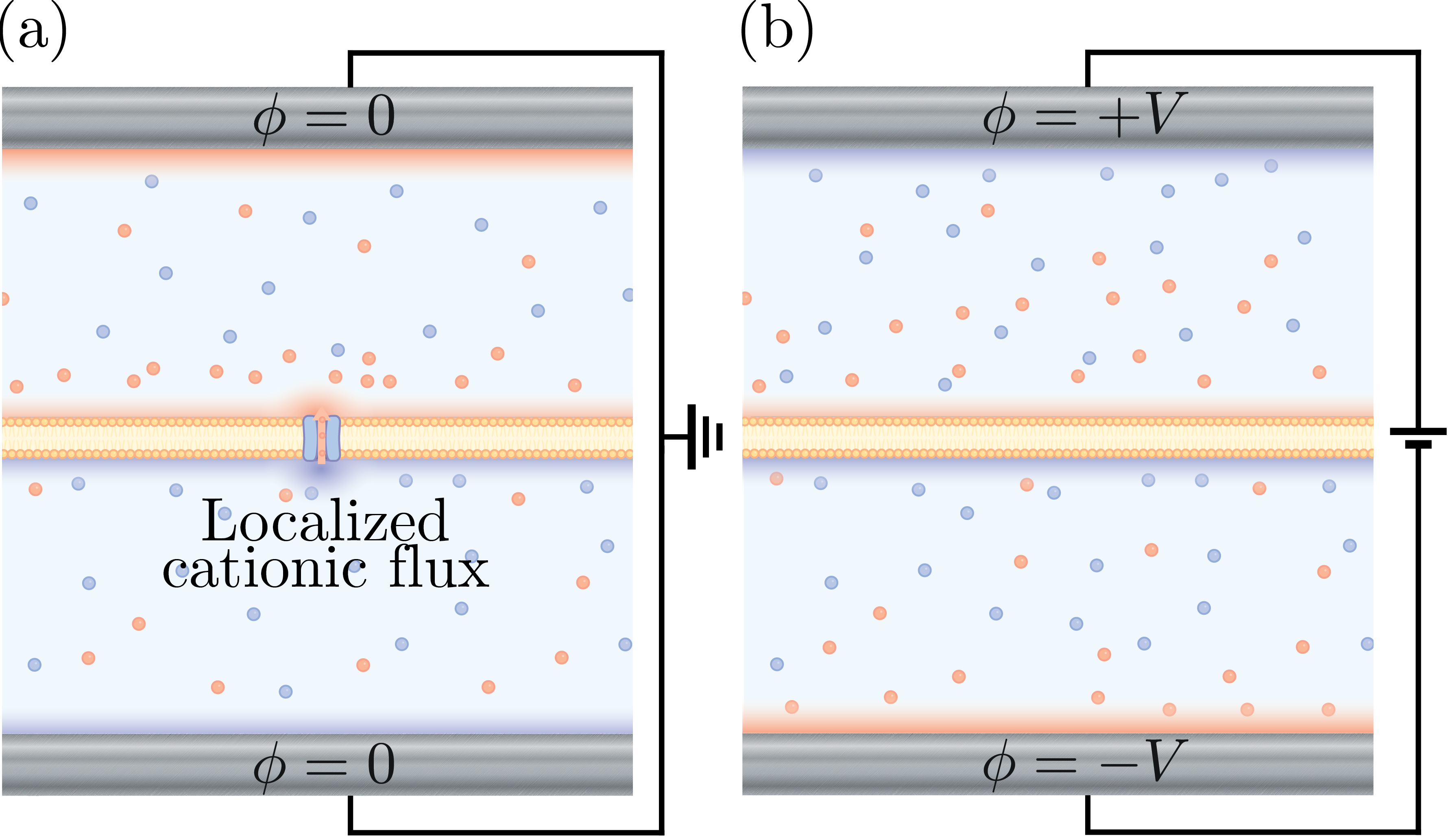}
    \caption{
    \textbf{Linear decomposition.} Under the linear approximation, we can decompose the overall problem depicted in Fig.~\ref{fig1_schematic_descriptions}  into two subproblems: (a) transmembrane current with zero applied voltage (C), analyzed here, and (b) impermeable membrane with a $2V$ potential difference across the electrodes (V), analyzed in Ref.~\cite{farhadi2025capacitive}
    }
\label{fig2_decomposition_of_problem}
\end{figure}

Third, linearity allows us to conveniently separate the contribution of the transmembrane ionic current from that of the applied voltage. This decomposition is illustrated schematically in Fig.~\ref{fig2_decomposition_of_problem}. Explicitly, we represent the electric potential and charge density as the superposition of two separate components
\begin{subequations}
\label{eq:superposition}
\begin{align}
    \phi &\eq \phi^{\rm C} + \phi^{\rm V} \ , \label{eq:superpositionPhi}\\
    \rho &\eq \rho^{\rm C} + \rho^{\rm V} \label{eq:superpositionRho}\ ,
\end{align}
\end{subequations}
where $\phi^{\rm C}$ and $\rho^{\rm C}$ correspond to the scenario with a transmembrane ionic current (${\rm C}$) but no applied voltage (Fig.~\ref{fig2_decomposition_of_problem}(a)), while $\phi^{\rm V}$ and $\rho^{\rm V}$ describe the complementary problem with an externally applied voltage  (${\rm V}$) but no transmembrane current (Fig.~\ref{fig2_decomposition_of_problem}(b)).

The voltage-driven subproblem has recently been solved in Ref.~\cite{farhadi2025capacitive}. Under the conditions ${\lambda_{\rm D}\ll L}$, ${\tau_{\rm D}\ll\tau_{\rm C}}$, and ${t\gg \tau_{\rm D}}$, the solutions are
\begin{subequations}
\label{eq:Jafar_sols}
\begin{align}
    \begin{split}\phi^{\rm V} &\eq V - V\left(1-\frac{z}{L}\right)e^{-\frac{t}{\tau_{\rm C}}}\\
    &\qquad - \frac{V}{2+\chi}
    \left(e^{-\bar{z}}-e^{-\bar{z}'}+1\right)\left(1-e^{-\frac{t}{\tau_{\rm C}}}\right)
    \end{split}\ , 
    \label{eq:Jafar_phi}\\
    \rho^{\rm V} &\eq \frac{V\epsilon}{\lambda_{\rm D}^2\left(2+\chi\right)}\left(e^{-\bar{z}}-e^{-\bar{z}'}\right)\left(1-e^{-\frac{t}{\tau_{\rm C}}}\right) \ ,
    \label{eq:Jafar_rho}
\end{align}
\end{subequations}
where $\bar{z} = z/\lambda_{\rm D}$ and $\bar{z}' = (L-z)/\lambda_{\rm D}$ are nondimensionalized axial coordinates, and $\chi$ and $\tau_{\rm C}$ are the capacitance ratio and the capacitive timescale, as defined previously. Due to the symmetry and in-plane uniformity of the problem (Fig.~\ref{fig2_decomposition_of_problem}(b)), the voltage-driven solution is independent of the radial coordinate $r$. 

For most of the article, we focus on the current-driven subproblem (C), with ${V=0}$, and verify the superposition in Sec.~\ref{subsec:Superposition}. To simplify notation, we omit the superscript C from now on.

\section{Boundary layer theory}
\label{sec:BL_Theory}

In principle, one strategy to analytically tackle Eqs.~\eqref{eq:dim_charge_bal_lin}-\eqref{eq:dim_eq_poisson_lin} would be to directly apply Laplace-Hankel transforms, as pursued in Ref.~\cite{row2025spatiotemporal}. In practice, however, the transformed solutions are difficult to invert exactly, and even asymptotic evaluations become unwieldy. Instead, we formulate a boundary layer theory that exploits the small parameter $ {\eta \equiv \lambda_{\rm D}/{L}\ll1}$, reflecting the extreme thinness of each diffuse charge layer relative to the macroscopic length $L$. By performing a perturbation analysis in $\eta$~\cite{Hinch_perturbation, bender_orszag},
we obtain compact leading-order equations that reveal the underlying physics. Crucially, the expansion shows that the bulk electrolyte behaves electroneutrally at all perturbational orders, and is coupled dynamically to the interfacial screening layers via time-dependent boundary conditions.  The remainder of this section summarizes the reduced equations of the boundary layer theory. Detailed derivations are provided in Sec.~III of the SM.

\subsection{Bulk electrolyte (Outer solutions)}
 Because of the sink term in Eq.~\eqref{eq:dim_charge_bal_lin}, any free charge in the bulk decays exponentially fast with the Debye timescale $\tau_{\rm D}$ and on the length scale $\lambda_{\rm D}$. Thus, for ${t\gg\tau_{\rm D}}$ and ${L\gg\lambda_{\rm D}}$, the bulk (denoted by superscript $\rm b$) solution remains electroneutral
\begin{equation}
    \rho^{\rm b} \eq 0 \ ,
\end{equation}
 at all perturbational orders. Accordingly, the bulk electric potential satisfies Laplace's equation 
\begin{equation}
    \nabla^{2} \phi^{\rm b}
    \eq
    0
    \ , \label{eq:dim_phi_reg_perturb_1st}
\end{equation}
with all free charges localized to the diffuse screening layers at membrane and electrode interfaces. We now determine the leading-order equations governing the distribution of charges and the electric potential in the diffuse layers, together with their coupling to the electroneutral bulk.

\subsection{Diffuse charge layers (Inner solutions)}
While the diffuse layers themselves are not electroneutral at leading order, their analysis can be simplified in two ways. First, the intrinsic timescale for local dynamics in these layers, $\tau_{\rm D}$, is much shorter than the characteristic charging times of the system $\tau_{\rm C}$ and $\tau_{\rm B}$ (Table~\ref{tab:parameters}). Thus, for ${t \gg \tau_{D}}$ the diffuse layers can be treated as quasistatic, and the time derivative in Eq.~\eqref{eq:dim_charge_bal_lin} can be dropped inside the layers.
Second, for ${r\gg \lambda_{\rm D}}$, axial gradients dominate over radial gradients inside the diffuse charge layer.
Therefore, we may omit the radial part of the Laplacian in Eqs.~\eqref{eq:dim_charge_bal_lin}-\eqref{eq:dim_eq_poisson_lin}, when applied near the membrane or electrode surfaces.

Consequently, the membrane-side boundary layer is described at leading order by
\begin{subequations}
\begin{align}
    -\frac{\partial^{2} \rho^{\rm m}}{\partial \bar{z}^{2}} + \rho^{\rm m}
    \eq
    0
    \ , \label{eq:dim_rho_sing_perturb_1st}
    \\
    -\frac{\epsilon}{\lambda_{\rm D}^{2}}\frac{\partial^{2} \phi^{\rm m}}{\partial \bar{z}^{2}}
    \eq
    \rho^{\rm m}
    \ , \label{eq:dim_phi_sing_perturb_1st}
\end{align}
\end{subequations}
where ${\bar{z} = z / \lambda_{\rm D}}$ is the \emph{stretched} axial coordinate within the membrane-side  (denoted by superscript m) diffuse layer, and ${\rho^{\rm m}(r, \bar{z}, t)}$ and $\phi^{\rm m}(r, \bar{z}, t)$ are the charge density and electric potential in this region. Analogous equations hold for ${\rho^{\rm e}(r, \bar{z}', t)}$ and ${\phi^{\rm e}(r, \bar{z}', t)}$ within the electrode-side (denoted by superscript e) diffuse layer, with a corresponding stretched axial coordinate ${\bar{z}' = (L - z) / \lambda_{\rm D}}$. From these equations, it is straightforward to obtain the quasistatic charge profiles inside the boundary layers as
\begin{subequations}
\label{eq:BL-charge}
\begin{align}
    \rho^{\rm m} (r, \bar{z}, t)
    &\eq
    \frac{\sigma^{\rm m}(r, t)}{\lambda_{\rm D}}  e^{-\bar{z}}
    \ , \label{eq:dim_soln_rho_sing_perturb_membrane}
    \\
    \rho^{\rm e} (r, \bar{z}', t)
    &\eq
    \frac{\sigma^{\rm e}(r, t)}{\lambda_{\rm D}}  e^{-\bar{z}'}
    \ , \label{eq:dim_soln_rho_sing_perturb_elec}
\end{align}
\end{subequations}
where ${\sigma^{\rm m}(r, t)}$ and ${\sigma^{\rm e}(r,t)}$ are the local areal charge densities at the membrane and electrode surfaces, respectively.

The exponential profiles show that the charge density decays to zero within a few Debye lengths in the axial direction to match the bulk value. Solutions for the electric potential inside the boundary layers in terms of $\sigma^{\rm m}$ and $\sigma^{\rm e}$ are presented in Sec.~III of the SM.

\subsection{Matching the bulk and diffuse charge layers}
To close the problem, we must relate the boundary layer quantities $\sigma^{\rm m}$ and $\sigma^{\rm e}$ to the bulk electric potential $\phi^{\rm b}$. To this end, we first asymptotically match the electric potential of the outer and inner regions ~\cite{Hinch_perturbation, bender_orszag}, which yields
\begin{subequations}
\label{eq:van_dyke_matching}
\begin{alignat}{3}
    \phi^{\rm b}&|_{z = 0} 
    &&\eq
    \lim_{\bar{z}\rightarrow\infty} \phi^{\rm m}
    &&\eq
    \frac{\sigma^{\rm m}(r,t)}{2C_{\rm M}^{\rm eq}}  
    \ , \label{eq:phi_matched_asymp_memb}
    \\
    \phi^{\rm b}&|_{z = L}
    &&\eq
    \lim_{\bar{z}'\rightarrow\infty} \phi^{\rm e}
    &&\eq
    \frac{\sigma^{\rm e}(r,t)}{C_{\rm D}}
    \ ,\label{eq:phi_matched_asymp_elec}
\end{alignat}
\end{subequations}
where ${C_{\rm M}^{\rm eq}\equiv \left[C_{\rm M}^{-1}+2 C_{\rm D}^{-1}\right]^{-1}=\epsilon/(2\lambda_{\rm D}(1+\chi))}$ is the equivalent capacitance of the membrane and its adjacent diffuse charge layers. The factor of 2 in the membrane-side equation arises since $2\phi^{\rm b}|_{z=0}$ describes the potential drop across the membrane and its two adjacent diffuse layers. This total drop is related to the transmembrane potential ${V^{\rm M}\equiv
2 \phi^{\rm m}(r, 0, t)}$ as
\begin{equation}
\label{eq:memb_pot_surf_chg}
V^{\rm M}(r,t)
\eq
\frac{\sigma^{\rm m}(r,t)}{C_{\rm M}}
\eq
\frac{C_{\rm M}^{\rm eq}}{C_{\rm M}} 2\phi^{\rm b}(r,0,t)  \ .
\end{equation}
Due to the low membrane capacitance (${\chi\gg1}$), we have ${C_{\rm M}^{\rm eq}}/{C_{\rm M}= \chi/(\chi+1) \approx 1}$, implying that ${V^{\rm M}(r,t)\approx2\phi^{\rm b}|_{z=0}}$.

Next, we match the charging dynamics of the inner regions with their leading-order driving force in the outer region, electromigration. This is done by integrating the full charge balance equation~\eqref{eq:dim_charge_bal_lin} across each boundary layer along the axial coordinate. Combined with Eqs.~\eqref{eq:van_dyke_matching}, this yields time-dependent boundary conditions for the bulk electric potential
\begin{subequations}
\label{eq:bl_bcs}
\begin{alignat}{2}
    &\left.
        2C_{\rm M}^{\rm eq}\frac{\partial \phi^{\rm b}}{\partial t}
    \right|_{z=0}
    &&\eq
    g \left.
        \frac{\partial \phi^{\rm b}}{\partial z}
    \right|_{z=0}
    +
     I
        \frac{\delta(r)}{2\pi r} 
    \ , \label{eq:phi_bulk_BC_memb}
    \\
    &\left.
        C_{\rm D}\frac{\partial \phi^{\rm b}}{\partial t}
    \right|_{z=L}
    &&\eq
    - g
    \left.
        \frac{\partial \phi^{\rm b}}{\partial z}
    \right|_{z=L}
    \ .\label{eq:phi_bulk_BC_elect}
\end{alignat}
\end{subequations}
In both equations above, the left-hand side describes the rate of charging the boundary layer, while the first  term on the right-hand side captures the current entering from the bulk via electromigration. In  Eq.~\eqref{eq:phi_bulk_BC_memb}, the final term represents the additional charge source due to the transmembrane current.
Solving Eqs.~\eqref{eq:dim_phi_reg_perturb_1st} and \eqref{eq:bl_bcs} gives the bulk potential $\phi^{\rm b}(r,z,t)$; the accompanying areal charge densities follow from Eqs.~\eqref{eq:van_dyke_matching}, and can be used to obtain the diffuse charge layer profiles in Eqs.~\eqref{eq:BL-charge}. 

From Eqs.~\eqref{eq:bl_bcs}, we identify two velocity scales, ${v \equiv g/2C_{\rm M}^{\rm eq}=(1 + \chi) D/ \lambda_{\rm D} }$ and ${u \equiv g/C_{\rm D}=D/\lambda_{\rm D}}$ that will recur throughout the remaining analysis. Each speed is the ratio of a conductivity to a capacitance.
The larger speed $v$ was previously identified as the in-plane propagation speed of the electrochemical signal separating the monopolar and dipolar regimes in an unbounded electrolyte~\cite{row2025spatiotemporal}. Both $u$ and $v$ can be recast in terms of familiar $RC$ timescales, highlighting their role in governing the dynamics of the system:
\begin{equation}
v
\eq
\left(\frac{1+\chi}{2+\chi}\right) \frac{L}{\tau_{\rm C}}
\ \approx \ 
\frac{L}{\tau_{\rm C}} \ , \quad
u
\eq
\frac{L}{\tau_{\rm B}} \ ,
\end{equation}
which links $v$ to the capacitive time scale $\tau_{\rm C}$ and $u$ to the $RC$ timescale of the bare electrolyte $\tau_{\rm B}$. As shown below, these two speeds demarcate the distinct temporal charging regimes for ${r>L}$.

\subsection{Summary and regime of validity}\label{subsec:bl_summary}
The boundary layer analysis reveals a coupled structure of charge transport. While the bulk electrolyte remains electroneutral, it supports electric fields that drive electromigration.
The ion-selective transmembrane current generates and maintains bulk electric fields, which continually transport charge into the quasistatic diffuse layers. The charge accumulates in the interfaces, and the resulting potential drops are governed by capacitive relations. In turn, the accumulation of these interfacial charges continuously reshapes the electric field in the bulk. 

As an additional remark, we note that the boundary layer theory relies on a clear separation of length and time scales, specifically ${\lambda_{\rm D}/L = \eta \ll 1}$ and ${\tau_{\rm D}/\tau_{\rm C} \sim \eta}$ \footnote{Since ${\tau_{\rm D}/\tau_{\rm C}= (n_{\rm D}/2 + \chi) \eta}$, this ratio might not be small for large $\chi$ even when ${\eta\ll 1}$. However, choosing any timescale asymptotically greater than $\tau_{\rm D}$ (e.g., $\tau_{\rm B}$) preserves the validity of the leading-order equations. See Sec.~III of the SM.}. The boundary layer theory is valid when the diffuse layers become quasistatic (${t>\tau_{\rm D}}$) and at radial locations where the axial transport dominates radial transport within the diffuse layers (${r>\lambda_{\rm D}}$).
To analyze early-time (${t \lesssim \tau_{\rm D}}$) and near-pore (${r \lesssim \lambda_{\rm D}}$) behaviors beyond the reach of boundary layer theory, we employ methods developed in Ref.~\cite{row2025spatiotemporal}. Specifically, for early times, we invoke an approximation that treats all the transported charges as localized point charges  on either side of the membrane, and solve the resulting electrostatics problem. Near the pore, the dynamics are governed by thermal diffusion and reaches a steady state within $\tau_{\rm D}$. In the following section, we show that these complementary approaches can be fruitfully combined with the boundary layer theory to understand the spatiotemporal dynamics of the system in all regimes.

\section{Results}
\label{sec:Results}
With the boundary-layer framework in place, we now present closed-form analytical expressions for the spatiotemporal behavior of $V^{\rm M}$. We confirm the accuracy of these expressions by comparing them with numerical solutions to the full PNP equations~\eqref{eq:nonlinear_eqs} and \eqref{eq:dim_eq_poissonM} while discussing their physical implications.
The numerical calculations use a nonlinear finite-element scheme~\cite{papadopoulos2015fem,hughes2012finite} following the meshing strategy of Ref.~\cite{row2025spatiotemporal}, implemented with \texttt{FEniCSx} \cite{Scroggs22a,Scroggs22b,Alnaes14} and \texttt{Gmsh} \cite{Geuzaine09,remacle2012blossom}.  Complete derivations are provided in Secs.~IV and~V of the SM. We organize the discussion of the results into the near-field (${r<L}$) behavior, covered in Sec.~\ref{subsec:NearField}, and the far-field (${r>L}$) behavior, covered in Sec.~\ref{subsec:FarField}. This division is motivated by our finding that the influence of electrodes on the near-field response is negligible until late times, whereas screening effects significantly impact the far-field behavior throughout.
\begin{figure}[t]
\centering
\includegraphics[width=\linewidth]{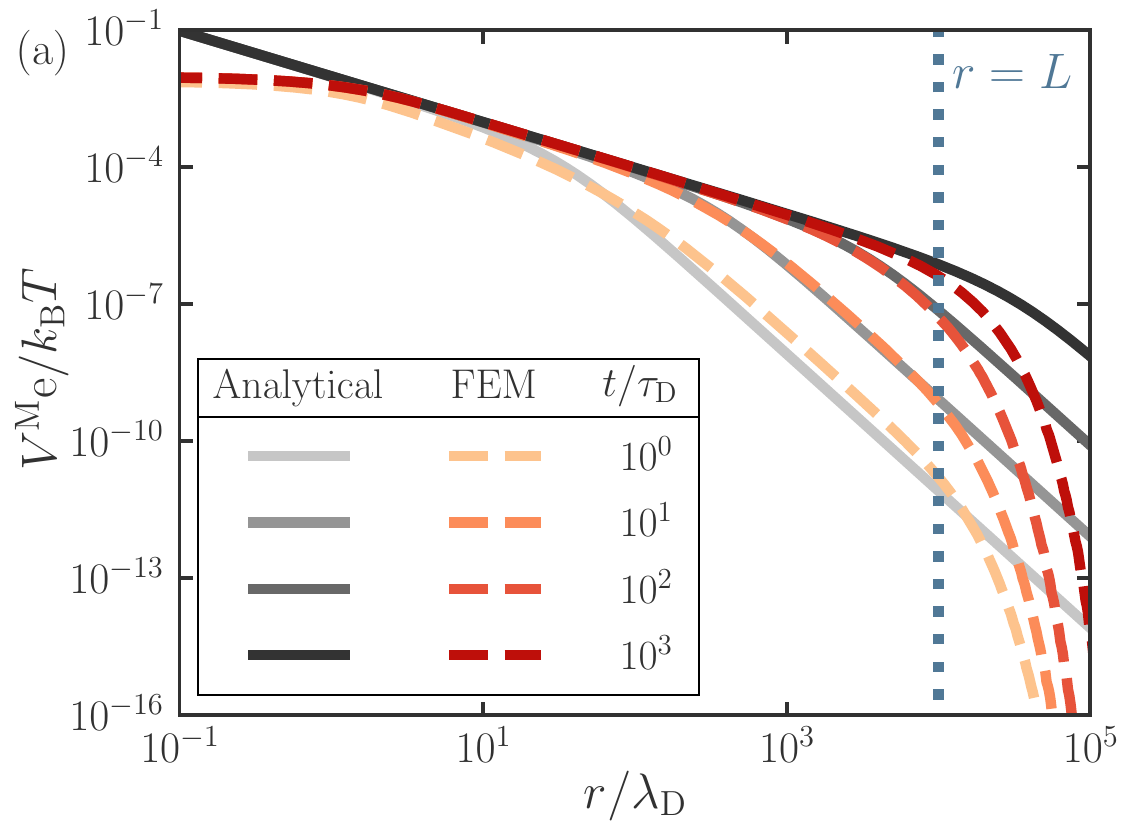}
\includegraphics[width=\linewidth]{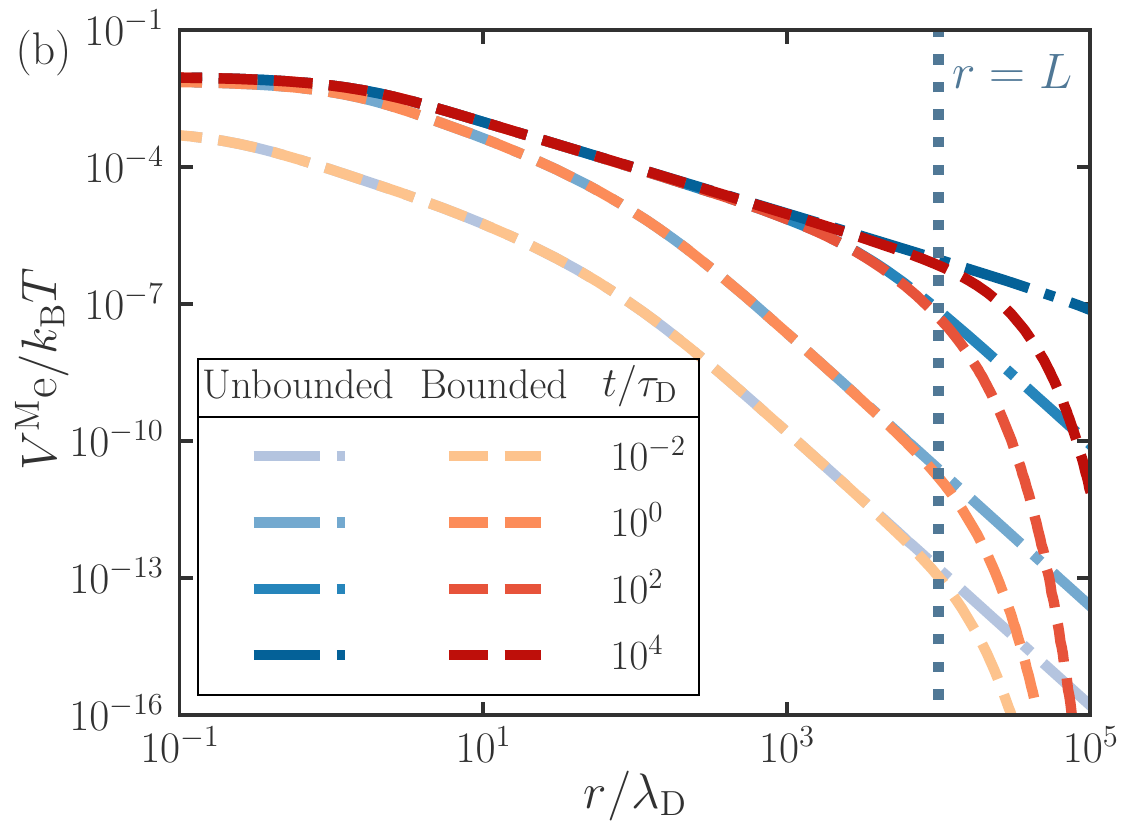}
    \caption{\textbf{Near-field behavior.} (a) Equation \eqref{eq:near-field} for $V^{\rm M}$ agrees with numerical simulations throughout the near field region. Agreement holds once the diffuse layers are charged (${t>\tau_{\rm D}}$) and within the radial band ${\lambda_{\rm D}<r<L}$.
    (b) The bounded-system simulations coincide with the results of Ref.~\cite{row2025spatiotemporal} for an unbounded electrolyte for ${t<\tau_{\rm B}}$ and ${r<L}$; deviations appear only when the radius exceeds the electrode gap (${r>L}$). On the y-axis, $V^{\rm M}$ is plotted in units of the thermal voltage ${k_{\rm B}T/{\rm e} = \text{25 mV}}$. Simulation parameters: ${\delta^{\rm M}=4\lambda_{\rm D}}$, ${\chi=40}$, ${I=0.06{\rm e}C_0 D\lambda_{\rm D}}$, ${R^{\rm P}=0.1\lambda_{\rm D}}$, ${L=10^4\lambda_{\rm D}}$.
    }
\label{fig4_like_unbounded_near_pore}
\end{figure}
\begin{figure}[t]
\centering
\includegraphics[width=\linewidth]{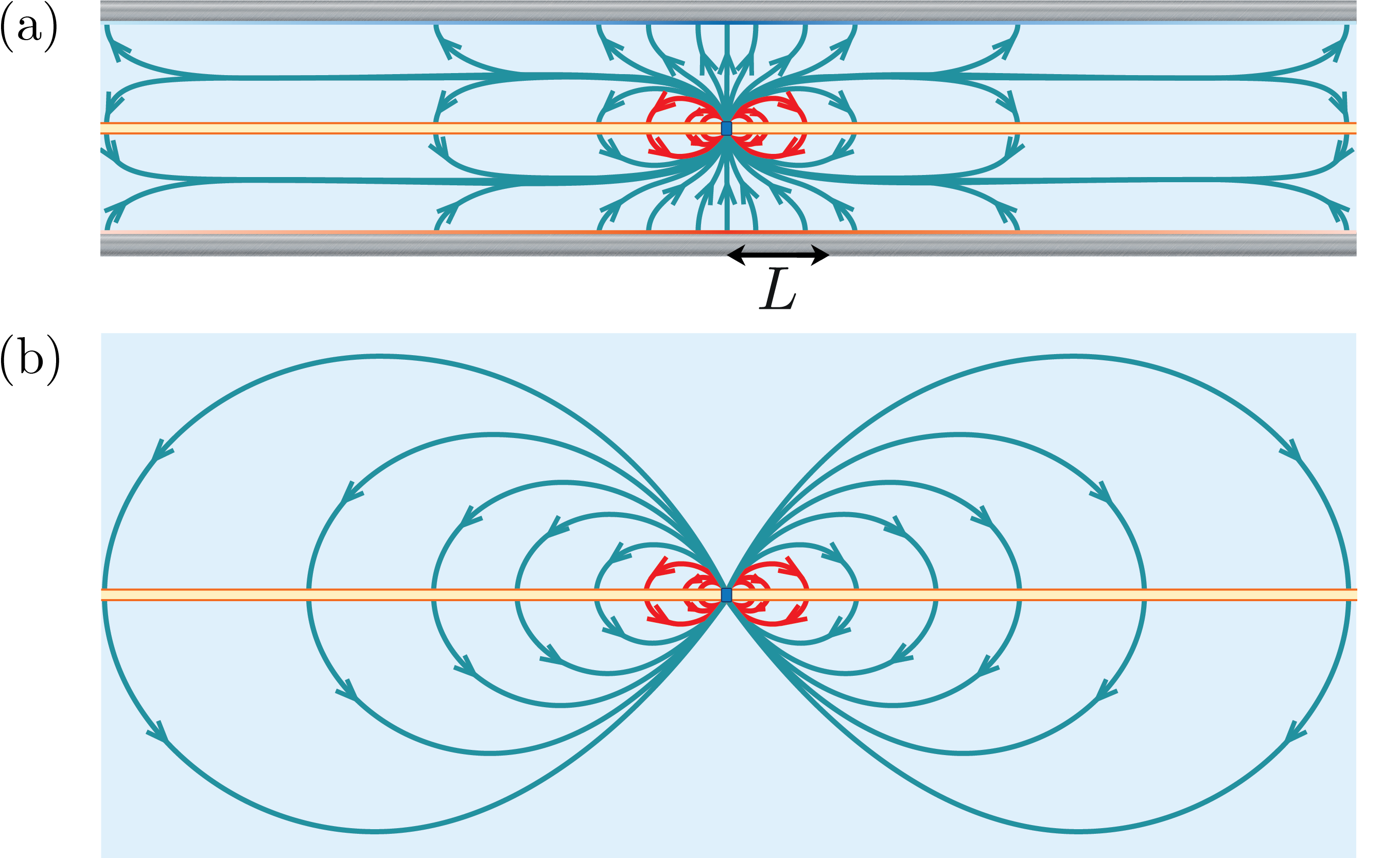}
    \caption{\textbf{Electric field structure at early times.} Schematic of electric-field lines based on the point-charge approximation, valid at early times for (a) an electrolyte bounded by two parallel electrodes (this paper), and (b) an unbounded electrolyte~\cite{row2025spatiotemporal}. The orange slab represents the membrane. In the bounded case, ion pumping promptly induces equal-and-opposite surface charge on the electrodes, significantly modifying the overall electric field structure. Yet, close to the transporter (${r < L}$), the electrodes are effectively invisible,  and the field pattern (red) is identical to the unbounded case. This invisibility is not permanent---at longer times confinement effects start to play a role, and the system enters a diffusive regime.
    }
\label{fig_electrostatics_E_field}
\end{figure}
\subsection{Near-field (${r<L}$) behavior matches \\ unbounded case}
\label{subsec:NearField}
In the limit $r\ll L$, the boundary layer analysis yields
\begin{equation}
\label{eq:near-field}
    V^{\rm M}(r,t)=\frac{I}{g\pi}\,
    \frac{\chi}{1+\chi}\!
    \left[\frac{1}{r}-\frac{1}{\sqrt{r^{2}+(vt)^{2}}}\right] ,
\end{equation}
which is identical to the profile of the transmembrane potential for an unbounded electrolyte \cite{row2025spatiotemporal}.  The solution contains a dynamic crossover radius $r^{*}(t)=vt$, where $v$ is the characteristic speed introduced previously. For $r<r^{*}$ the potential shows a monopolar $1/r$ decay, while for $r>r^{*}$ it switches to a dipolar $1/r^{3}$ form. 
Figure~\ref{fig4_like_unbounded_near_pore}(a) confirms that Eq.~\eqref{eq:near-field} matches the numerical simulations for ${\lambda_{\rm D} < r \lesssim L}$ once the initial Debye-layer charging is complete (${t>\tau_{\rm D}}$).  Deviations at ${r\lesssim \lambda_{\rm D}}$ or ${t\lesssim \tau_{\rm D}}$ arise because the boundary layer analysis breaks down in this regime; however, the dynamics in this regime were resolved in our previous work \footnote{See Section IV.2 in Supplemental Material of Ref.~\cite{row2025spatiotemporal}}.
Figure~\ref{fig4_like_unbounded_near_pore}(b) shows that the bounded- and the unbounded-case profiles overlap for ${r < L}$, suggesting that the electrodes have no discernible effects in this zone---at least until the timescale $\tau_{\rm B}$, as we shall see later.

Why is this the case? The electric field lines at early times ${t<\tau_{\rm D}}$ (Fig.~\ref{fig_electrostatics_E_field}) provide an intuitive explanation. Field lines emanating from the outlet of the pump must either loop back to the membrane or terminate on the electrode surface. Near the pump (${r < L}$), the field lines are minimally perturbed by the presence of the electrodes, and the diffuse charge dynamics in these regions can be described by the unbounded-domain theory. This agreement persists even beyond the early-time regime, since the evolution of the charge density remains governed by the electric field, which itself is largely unaffected by the electrode boundaries for ${r < L}$.
However, this correspondence is not permanent. At late times ${t>\tau_{\rm B}}$, confinement effects become significant and the charge accumulated at ${r > L}$ distorts the electric fields within ${r < L}$. As a result, the mechanism of charge reorganization in ${r < L}$ changes, giving rise to new behavior, as discussed in the next section.

\begin{figure*}
    \centering
    \includegraphics[width=\linewidth]{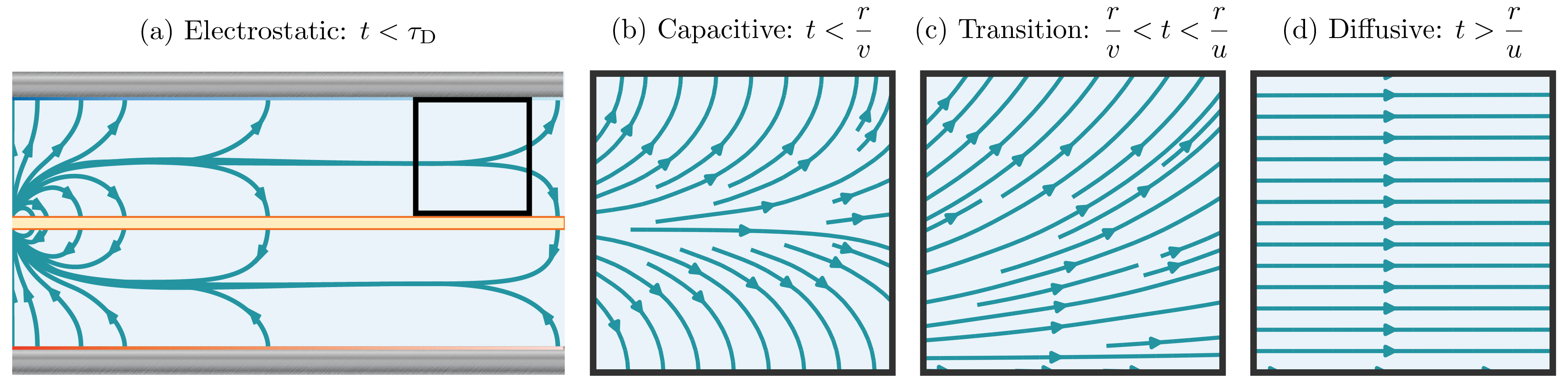}
    \caption{\textbf{
    Snapshots of electric field evolution.}
    Electric field lines in the far-field (inset for ${10L<r<11L}$, ${0<z<L}$ as shown in (a)) at four times: (a) ${t<\tau_{\rm D}}$ (electrostatic regime as in Fig.~\ref{fig_electrostatics_E_field}(a)), (b) ${t=10^{-2}\tau_{\rm C}}$ (capacitive regime), (c) ${t=10\tau_{\rm C}}$ (transition regime), and (d) ${t=10^2\tau_{\rm B}}$ (diffusive regime). Initially, panels (a) and (b) show a field symmetric about ${z=L/2}$ as in Fig.~\ref{fig_electrostatics_E_field}(a), and diffuse charge accumulates symmetrically at the membrane and electrode surfaces. However, because the membrane’s capacitance is much smaller than that of the electrode-side diffuse charge layer, a net axial field develops, as in panel (c), driving additional charge toward the electrode during the capacitive and transitional regimes. For ${t> r/u}$, axial variations disappear and only a radial field remains, panel (d), signaling the onset of the 2D diffusive regime.  Simulation parameters as in Fig.~\ref{fig4_like_unbounded_near_pore} but with ${L=10^3\lambda_{\rm D}}$.
    } 
    \label{fig5_efields}
\end{figure*}

\subsection{Far-field  $({r>L})$ behavior shaped by electrode-mediated screening effects}
\label{subsec:FarField}
Figure~\ref{fig4_like_unbounded_near_pore}(b) reveals a clear departure between the profiles of $V^{\rm M}$ of the bounded and unbounded cases when the radial distance exceeds the electrode-membrane gap, i.e. ${r>L}$. To understand this far-field behaviour, we break the dynamics into well-defined time windows.  For ${t\lesssim\tau_{\rm D}}$, we treat the transported charge as a point source, following Ref.~\cite{row2025spatiotemporal} (Sec.~\ref{sec:early_time}). Beyond the Debye time (${t>\tau_{\rm D}}$), the diffuse layers become quasistatic, and a boundary-layer analysis reveals three successive response regimes punctuated by the capacitive timescale $\tau_{\rm C}$ and the bare electrolyte $RC$ timescale $\tau_{\rm B}$; these regimes are detailed in Secs.~\ref{sec:capacitive}-\ref{sec:diffusive}.

\subsubsection{Electrostatic regime (${t\lesssim\tau_{\rm D}}$)}
\label{sec:early_time}
When ${t \lesssim \tau_{\rm D}}$, screening layers are yet to form and the diffuse charge remains essentially immobile. We therefore treat the membrane and surrounding electrolytes as ideal dielectrics and represent the net charge transferred or left behind as point charges.
Solving the resulting electrostatics problem for ${r\gg L}$ yields 
\begin{equation}
\label{eq:point_chg_soln}
    V^{\rm M}_{\rm pc}(r, t) \eq \frac{2\pi I t}{\epsilon L}\left(\frac{\chi \lambda_{\rm D}}{L}\right)^2 K_0\left(\frac{\pi r}{L}\right) \ ,
\end{equation}
where $I t$ is the total transported charge and $K_0$ is the zeroth-order modified Bessel function of the second kind. For ${x\gg 1}$, ${K_0(x)\sim e^{-x}/\sqrt{x}}$, so the transmembrane potential decays exponentially in-plane with a characteristic length $L/\pi$.

The exponential decay is a direct consequence of electrode-induced screening. In an unbounded electrolyte the transmembrane potential created by an active ion transporter decays as a power law in $r$ \cite{row2025spatiotemporal}, reflecting long-ranged charge electrostatic reorganization. Introducing electrodes confines the long-ranged character of the disturbance to a radial distance ${r\sim L}$; beyond this distance, the signal attenuates exponentially. Physically, charge transported across the membrane immediately induces surface charge on the electrodes. At large distances, the net effect of the transported charge and induced surface charge is to screen the response on the length scale of the electrode spacing (see Fig.~\ref{fig_electrostatics_E_field}(a)).

\subsubsection{Capacitive regime (${\tau_{\rm D}<t < r/v}$)}
\label{sec:capacitive}
Once the Debye time has lapsed (${t>\tau_{\rm D}}$) the diffuse layers have equilibrated locally, so the boundary-layer theory becomes applicable. Recall that the time-dependent boundary conditions in Eqs.~\eqref{eq:bl_bcs} involve the characteristic speeds, ${v \approx L/\tau_{\rm C}}$ and ${u \approx L/\tau_{\rm B}}$. In the unbounded problem, a single transition at ${t=r/v}$ separates the dipolar from the monopolar behavior at any location $r$. With electrodes present the dynamics at a fixed ${r>L}$ instead unfold in three successive windows of time: (a) ${t<r/v}$, (b) ${r/v<t<r/u}$, and (c) ${t>r/u}$. We label these the \emph{capacitive}, \emph{transitional}, and \emph{diffusive} regimes, respectively. Consistent with this tripartite structure, field-line snapshots from the simulations (Figs.~\ref{fig5_efields}(a)-(d)) show different patterns in each window. We analyze the capacitive and diffusive regimes and describe the transitional regime as a crossover between the two limiting cases.

\begin{figure}[!tp]
\centering
\includegraphics[width=0.97\linewidth]{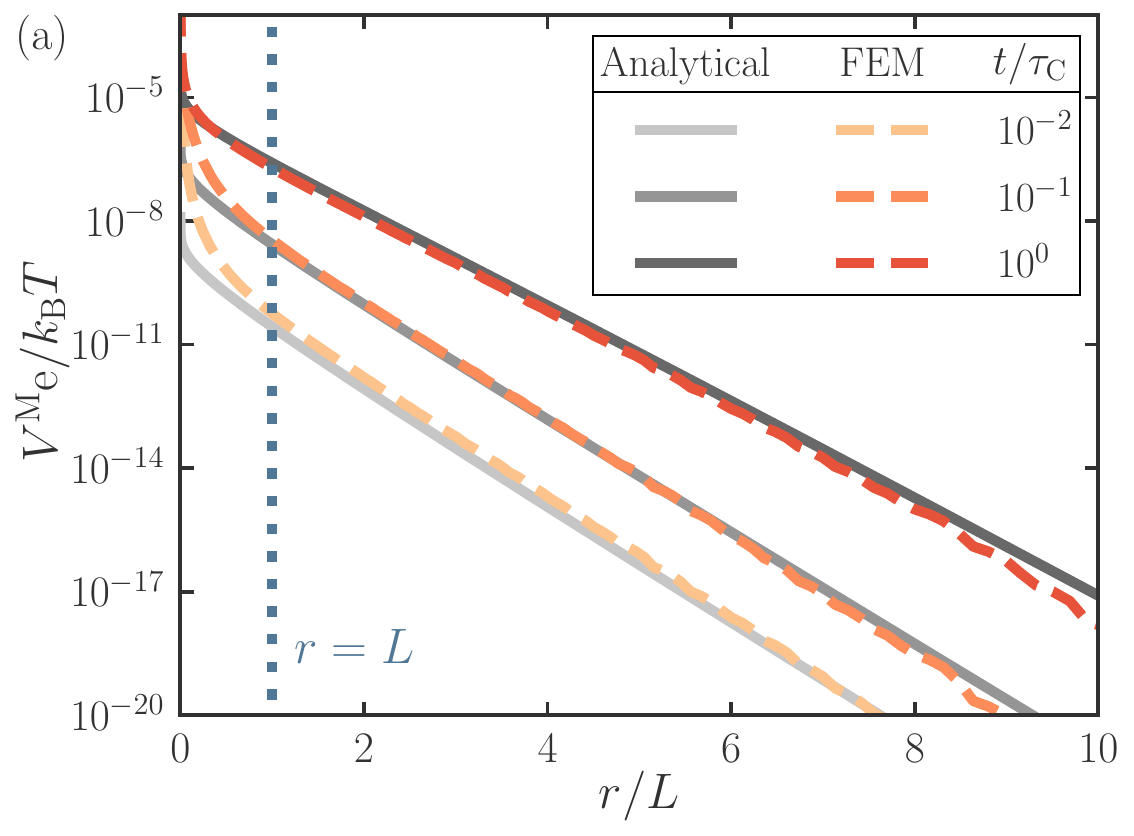}
\includegraphics[width=0.97\linewidth]{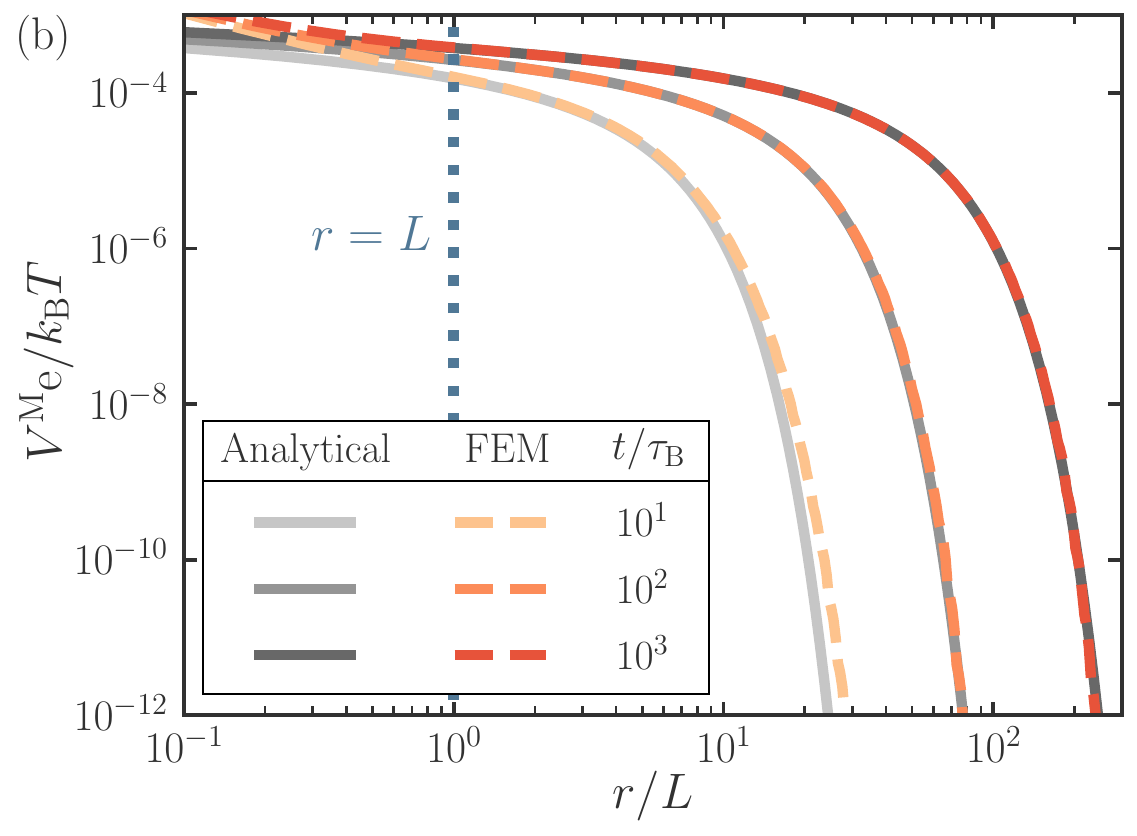}
\includegraphics[width=0.97\linewidth]{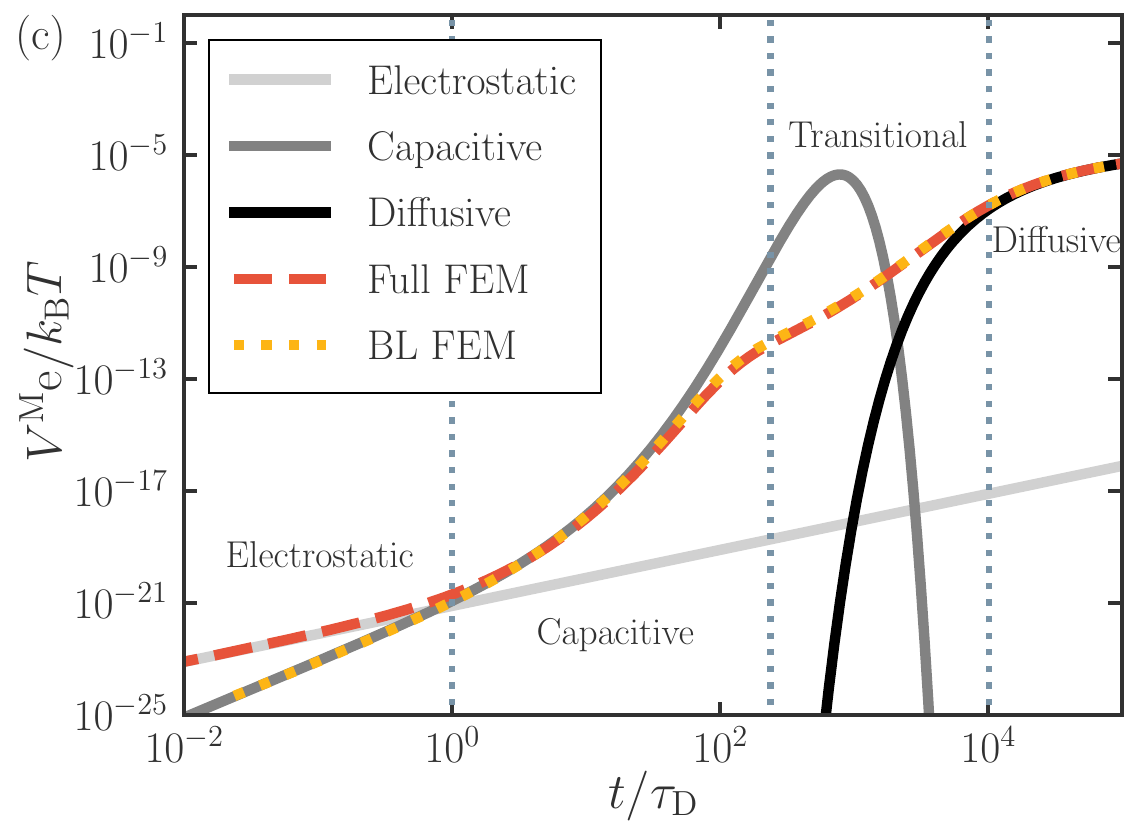}
    \caption{\textbf{Far-field behavior.} Analytical solutions of $V^{\rm M}$ for ${r>L}$ compared with numerical results for (a) the capacitive regime, showing Eq.~\eqref{eq:early_time} matches simulations when ${t/\tau_{\rm C}<r/L}$, (b) the diffusive regime, showing Eq.~\eqref{eq:late_time} matches simulations for ${t/\tau_{\rm B}>r/L}$. (c) $V^{\rm M}$ at a fixed radial distance ${r=10L}$ versus time, illustrating seamless agreement with the three asymptotic formulas across their respective domains. The analysis does not cover the late-capacitive-to-transition interval that lies between the two regimes. Simulation parameters other than $L$ as in Fig.~\ref{fig4_like_unbounded_near_pore} for all simulations. Different system sizes were used in each simulation to provide sufficient separation of time scales: (a) ${L=10^4\lambda_{\rm D}}$; (b) ${L=10^2\lambda_{\rm D}}$; (c) ${L=10^3\lambda_{\rm D}}$.}
    \vspace{-0.1in}
\label{fig5_far_field}
\end{figure}

For the capacitive regime (${t<r/v}$), the leading-order boundary layer solution is
\begin{equation}
    V^{\rm M} \eq \frac{2I}{ g r}\frac{\chi(\chi+1)}{(\chi+2)^2}\frac{t}{\tau_{\rm C}}K_0\left(\frac{\pi r}{L}\right)I_2\left(2\sqrt{\frac{t}{\tau_{\rm C}}\frac{\pi r}{L}}\right)
    e^{-\frac{t}{\tau_{\rm C}}}
    \ , \label{eq:early_time}
\end{equation}
where $I_\nu$ is the modified Bessel function of the first kind. The factor ${K_0(\pi r / L)}$ carries over the exponential screening imposed by the electrodes. 
Because ${v \equiv L/\tau_{\rm C}}$, the conditions for the capacitive regime can also be expressed as ${t/\tau_{\rm C} < r / L}$. Figure~\ref{fig5_far_field}(a) shows agreement between Eq.~\eqref{eq:early_time} and the numerical simulations for three values of ${t\leq \tau_{\rm C}}$ satisfying this condition.

Examining the argument of $I_2$ in Eq.~\eqref{eq:early_time}, we may consider asymptotic limits comparing ${t/\tau_{\rm C}}$ with ${L/r}$. At very early portions of the interval ${t/\tau_{\rm C} < L/\pi r}$, the argument is small and  ${I_{2}(x)\approx x^{2}/8}$. In this limit
\begin{equation}
\label{eq:quad_potential}
    V^{\rm M}(r,t) \eq \frac{\pi I t^2}{\epsilon L\tau_{\rm D}}\left(\frac{
\chi \lambda_{\rm D}}{L}\right)^2 K_0\left(\frac{\pi r}{L}\right) \ ,
\end{equation}
where we also assume ${\chi\gg 1}$. Notably,  Eq.~\eqref{eq:quad_potential} resembles the early-time response in Eq.~\eqref{eq:point_chg_soln} but grows quadratically with time rather than linearly. The origin of the quadratic law becomes clear when one integrates the membrane surface charge balance Eq.~\eqref{eq:phi_bulk_BC_memb} over time for ${r>\lambda_{\rm D}}$ with application of Eq.~\eqref{eq:memb_pot_surf_chg}:
\begin{equation}
    C^{\rm M}V^{\rm M}(r,t)
    \eq
    g\int_0^t  
    \left.
    \frac{\partial \phi^{\rm b}}{\partial z}\right|_{z=0}
    \, d\tau 
    \ .
\end{equation}
The integrand ${\left.\partial \phi^{\rm b}/\partial z\right|_{z=0}}$ can be estimated from the potential drop across the membrane-side boundary layer. The areal charge density ${C^{\rm M}V^{\rm M}}$ gives rise to a potential drop ${(C^{\rm M}/C^{\rm D})V^{\rm M}}$ within the diffuse layer of width $\lambda_{\rm D}$, resulting in $\left.\partial \phi^{\rm b}/\partial z\right|_{z=0}\approx V^{\rm M}/(2\chi\lambda_{\rm D})$.
At times ${t/\tau_{\rm C} < L/\pi r}$, the diffuse charge is still negligible and $V^{\rm M}$ can be approximated with $V_{\rm pc}^{\rm M}$, resulting in
\begin{equation}
    V^{\rm M}(r,t)=\frac{1}{\tau_{\rm D}}\int_0^t V_{\rm pc}^{\rm M}(r,\tau) \, d\tau \ .
\end{equation}
Since $V_{\rm pc}^{\rm M}$ grows linearly with time, we expect $V^{\rm M}$ to have a quadratic time-dependence as in Eq.~\eqref{eq:quad_potential}.
In fact, the same mechanism underlies the quadratic behavior of the dipolar solution encapsulated in Eq.~\eqref{eq:near-field} under the limit ${vt\ll r}$.

Eventually, however, the accumulated surface charge alters the field enough that the point-charge approximation breaks down, and the rise in $V^{\rm M}$ is no longer quadratic in time. As time advances and ${t/\tau_{\rm C}>L/\pi r}$, but ${t < r/v}$ still, the large-argument form of $I_2$ in Eq.~\eqref{eq:early_time} gives
\begin{equation}
    V^{\rm M}(r,t) \sim \exp\left[
    -\frac{\pi r}{L}\left(1-\sqrt{\frac{vt}{\pi r}}\right)^2
    \right] \ ,
\label{eq:V^M_exp}    
\end{equation}
where we neglect all sub-exponential terms.  Thus, following the initial quadratic growth phase, the potential increases exponentially in $\sqrt{t}$ at leading order until ${t\sim r/v}$. On the other hand, the leading order spatial profile decays exponentially with a characteristic scale of ${L/\pi}$ throughout the capacitive regime. The comparison between analytical expressions and numerical simulations in Fig.~\ref{fig5_far_field}(c) captures these features.

Finally, as ${t\sim r/v}$, Eq.~\eqref{eq:early_time} begins to diverge from observed behavior, signaling the onset of a transitional regime. Here, the membrane potential continues to increase, before eventually entering the diffusive regime described in the next section.

\subsubsection{Diffusive regime (${t > r/u}$)}
\label{sec:diffusive}
\begin{figure}[!t]
\centering
\includegraphics[width=0.95\linewidth]{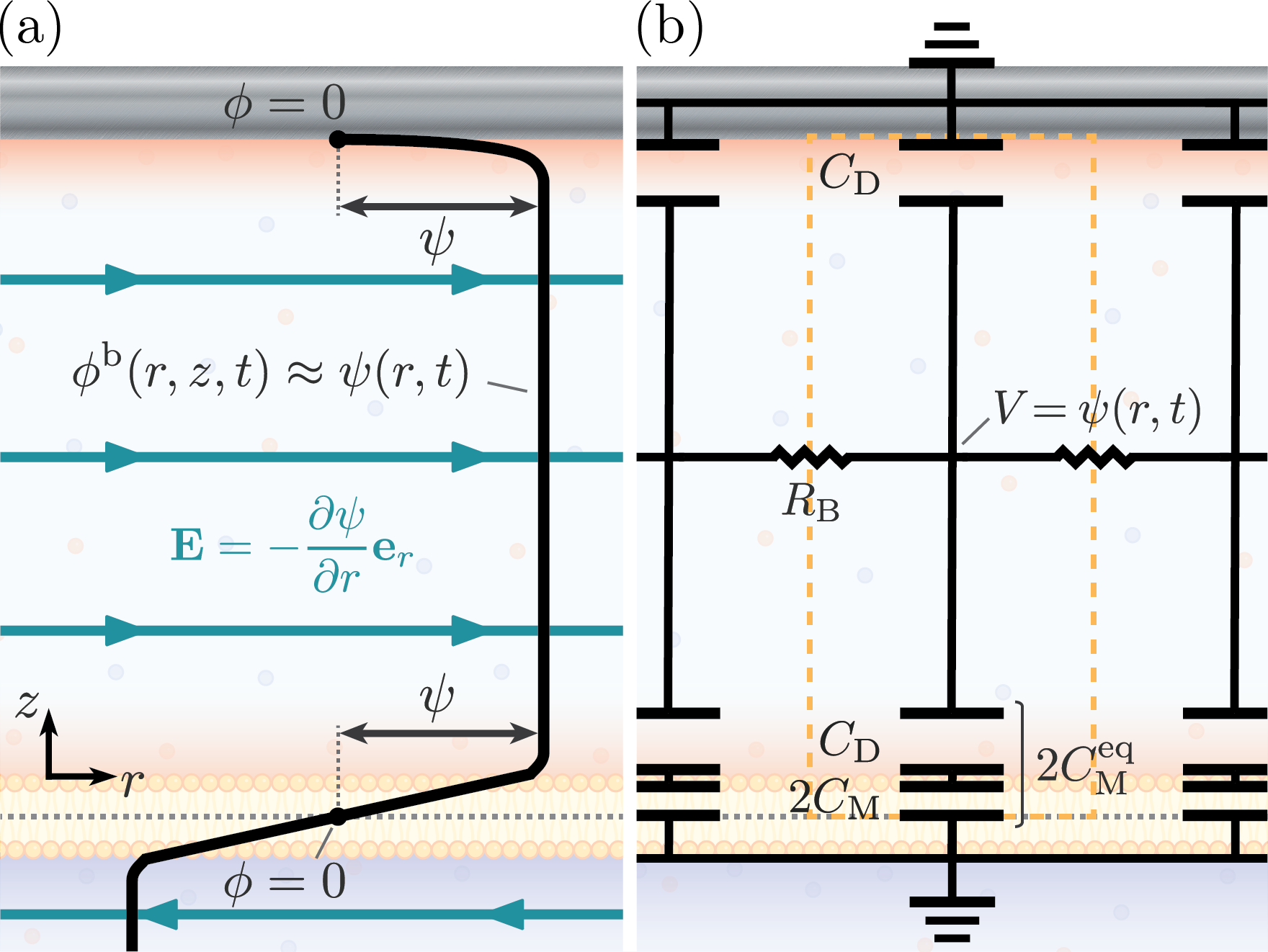}
    \caption{
    \textbf{Diffusive regime.} (a) At times ${t > r/u}$, the potential (black curve) becomes nearly uniform along $z$ in the bulk, where it is well-approximated by its axial average ${\phi^{\rm b}(r,z,t) \approx \psi(r,t)}$. The resulting electric field (teal lines) drives radial electromigration through the bulk.
    (b) An equivalent circuit picture of the diffusive behavior.  Each cylindrical slice of width $\Delta r$ (dashed box) is represented by capacitors corresponding to the diffuse charge layers and membrane. The electroneutral bulk supports a radial current between the slices with a resistance ${R_{\rm B} = \Delta r / (2 \pi r  L g)}$ while charge accumulates in the diffuse layers with areal charge densities ${C_{\rm D}\psi}$ (electrode-side) and ${2C^{\rm eq}_{\rm M}\psi}$ (membrane-side). Multiplying by $2\pi r\Delta r$ gives the total charge in the slice, which is therefore proportional to $\psi$. The radial charge flux is proportional to ${\partial \psi / \partial r}$ in limit ${\Delta r \rightarrow 0}$, leading to diffusive dynamics for the potential.
    }
    \vspace{-0.1in}
\label{fig_BL_diffusion}
\end{figure}
For times ${t>r/u}$, the bulk potential is nearly constant in $z$, as indicated by the field lines in Fig.~\ref{fig5_efields}(d). This motivates a dimension reduction of the axial direction. We average the bulk potential over $z$ by defining
\begin{equation}
    \psi(r, t) \eq \frac{1}{L}\int_0^L \phi^{\rm b}(r, z, t) \, dz \ .
\end{equation}
Integrating Eq.~\eqref{eq:dim_phi_reg_perturb_1st} across the membrane-electrode gap with the boundary conditions in Eqs.~\eqref{eq:bl_bcs}, and applying ${\phi^{\rm b}(r, z, t) \approx \psi(r,t)}$ yields a diffusion equation
\begin{equation}
    \label{eq:diff_eq_eff}
    (2C_{\rm M}^{\rm eq}+C_{\rm D})\frac{\partial \psi}{\partial t}=gL\frac{1}{r}\frac{\partial}{\partial r}\left(r \frac{\partial \psi}{\partial r}\right) + I\frac{\delta(r)}{2\pi r} 
    \ ,
\end{equation}
with an effective diffusivity
\begin{equation}
\label{eq:eff_diff}
    D_{\rm eff} \equiv \frac{gL}{2C_{\rm M}^{\rm eq}+C_{\rm D}}\eq 
    \frac{1+\chi}{2+\chi} \frac{L}{\lambda_{\rm D}}
    D \ .
\end{equation}

Equation~\eqref{eq:diff_eq_eff} represents an overall charge balance across a cylindrical slice of the system, as illustrated in Fig.~\ref{fig_BL_diffusion}(a). The left-hand term describes the rate of increase of total charge at $r$ stored in the diffuse layers adjacent to the membrane and electrodes, both proportional to the average bulk potential $\psi$. The dominant contribution to the radial flux comes from electromigration in the bulk and scales with the radial electric field, ${-\partial\psi/\partial r}$. Together with the localized current source from the ion transporter, this yields a diffusion equation for $\psi$.
The factor $L/\lambda_{\rm D}$ in $D_{\rm eff}$ reflects the geometry: electromigration acts across the full gap $L$, whereas free charges reside only in the diffuse charge layers of thickness $\lambda_{\rm D}$.
This charge transport process can be represented by an equivalent circuit, shown in Fig.~\ref{fig_BL_diffusion}(b),  in which capacitors represent charge storage at the membrane and electrodes, and resistors represent bulk migration. The resulting circuit is analogous to the classical cable model~\cite{dayan2001,rall1962} but includes an additional capacitive element representing the electrode-side diffuse layers.

Expressing the transmembrane potential as ${V^{\rm M} = (2C_{\rm M}^{\rm eq}/C_{\rm M})\psi}$ based on Eq.~\eqref{eq:memb_pot_surf_chg}, we recover a cable-like equation for the radial propagation of $V^{\rm M}$. The resulting equation can be solved to yield
\begin{equation}
\label{eq:late_time}
    V^{\rm M}(r,t) \eq \frac{I}{2\pi g L}\frac{\chi}{\chi+1}\left[-\operatorname{Ei}\left(-\frac{r^2}{4D_{\rm eff}t}\right)\right] \ ,
\end{equation}
where the exponential integral $\operatorname{Ei}$ comes from time-integrating the 2D diffusion kernel.
Figure~\ref{fig5_far_field}(b) compares Eq.~\eqref{eq:late_time} with fully nonlinear simulations at ${t/\tau_{\rm B}=10^1,10^2,10^3}$. As in the capacitive regime, the condition ${t>r/u}$ may be rewritten as ${r/L<t/\tau_{\rm B}}$. Accordingly, we see that for ${t=10\tau_{\rm B}}$, the diffusion solution holds up to ${r=10L}$. At later times, the diffusion solution holds for the entire plotted range, as expected.

For long times ${t> r^2/4D_{\rm eff}}$, Eq.~\eqref{eq:late_time} predicts that ${V^{\rm M}\propto \log\left(4D_{\rm eff}t/r^2\right)}$, suggesting that at any location the transmembrane potential grows logarithmically with continuous pumping. Thus, unlike the unbounded system---where the potential eventually approaches a monopolar steady-state profile---the long-time dynamics in the presence of blocking electrodes remain unsteady at all locations. At sufficiently large $t$, the diffusive contribution overtakes the near-field monopolar term even at distances ${r<L}$; the crossover occurs when
\begin{equation}
    \frac{t}{\tau_{\rm B}} \eq \left(\frac{r}{L}\right)^2\exp\left(\frac{L}{r}-1\right) \ ,
\label{eq:diffusion_onset}    
\end{equation}
so as ${r/L}$ decreases, the time required increases exponentially.
Thus, the diffusive regime first emerges at ${r = L}$ when ${t = \tau_{\rm B}}$ and subsequently spreads both inward and outward along the radial direction.

The time evolution of $V^{\rm M}$ at a fixed position ${r=10L}$ is presented in Fig.~\ref{fig5_far_field}(c).  The four temporal windows---electrostatic, capacitive, transitional, diffusive---are highlighted and the analytical solutions obtained by the point-charge approximation (${t<\tau_{\rm D}}$) and boundary layer theory (${t>\tau_{\rm D}}$) are compared with the numerical solutions to the full PNP equations. The analytical solutions accurately capture the dynamics within their respective regimes and the transitional regime appears as a smooth crossover from capacitive to diffusive behavior.

Figure~\ref{fig5_far_field}(c) also demonstrates agreement betwen the numerical solutions of the boundary layer theory (Eqs.~\eqref{eq:dim_phi_reg_perturb_1st} and~\eqref{eq:bl_bcs}) and the full PNP equations (Eqs.~\eqref{eq:nonlinear_eqs} and~\eqref{eq:dim_eq_poissonM}) for ${t > \tau_{\rm D}}$. This agreement affirms the computational utility of the boundary layer theory for ${t > \tau_{\rm D}}$ and ${r > \lambda_{\rm D}}$. In contrast to the full PNP equations, which require resolving steep spatial gradients across microscopic diffuse layers, and capturing fast dynamics over the microscopic Debye time, the boundary layer theory operates entirely at macroscopic spatial and temporal scales.  This enables the use of coarser meshes and larger time steps, leading to orders-of-magnitude computational speedup while maintaining comparable accuracy within its domain of validity. Consequently, the boundary layer theory offers a powerful and efficient framework especially for studying long-time dynamics, particularly in systems with complex geometries where the full PNP simulations may be computationally prohibitive.

\subsection{Superposed response to an applied voltage and a transmembrane current}
\label{subsec:Superposition}

\begin{figure}[t]
    \centering
    \includegraphics[width=\linewidth]{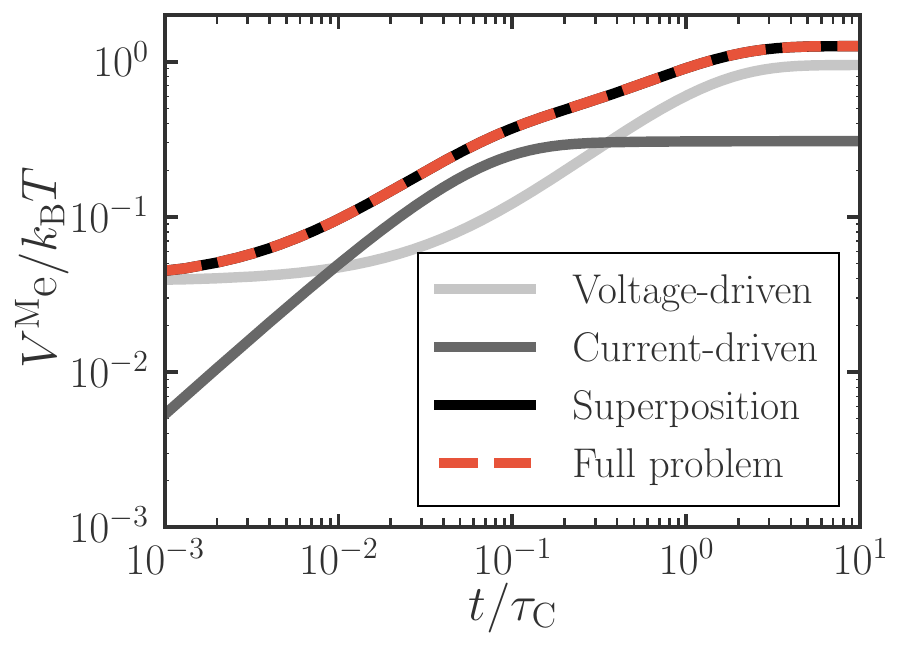}
    \caption{\textbf{Superposition of the current- and voltage-driven subproblems.} Transmembrane potential $V^{\rm M}$ at ${r=5\lambda_{\rm D}}$ as a function of time. The response to a simultaneous step current ${I=10 {\rm e}C^0D\lambda_{\rm D}}$ and a step voltage ${2V=k_{\rm B}T/{\rm e}}$ (red dashed line) coincides with the sum (black) of the voltage-driven (light gray) and current-driven (dark gray) subproblems. Other system parameters: ${\delta^{\rm M}=4\lambda_{\rm D}}$, ${\chi=40}$, ${R^{\rm P}=0.1\lambda_{\rm D}}$, ${L=10^3\lambda_{\rm D}}$.}
    \label{fig_superposition}
\end{figure}

The results in Sec.~\ref{subsec:NearField} and Sec.~\ref{subsec:FarField} are for the current-driven subproblem where both electrodes are maintained at zero potential. For sufficiently small applied potentials and transmembrane currents relevant for physiological settings, we can obtain the full response to the \emph{simultaneous} imposition of a step voltage $2V$ and a step transmembrane current $I$ through a superposition, as in Eq.~\eqref{eq:superposition}, of the current-driven and voltage-driven subproblems.
The solution to the voltage-driven subproblem is provided in Eq.~\eqref{eq:Jafar_sols}.
Figure~\ref{fig_superposition} confirms this expectation, where the full simulation with both stimuli (dashed red) agrees with the sum (solid black) of the voltage-only and current-only solutions (solid gray).

One may note that the superposition principle also extends to cases in which the membrane current depends on the applied voltage---for example, currents mediated by voltage-gated ion channels.  Provided the voltage remains within the linear regime, one may first prescribe the $I(V)$ relation from the channel model and then form the composite solution 
\begin{multline}
    \phi(r,z,t) = \int_{0}^{t}
    \bigg[ 
         \phi^{\rm V}(r,z,t-\tau)\frac{dV(\tau)}{d\tau} \\
        + \phi^{\rm C}(r,z,t-\tau)\frac{dI(\tau)}{d\tau} 
    \bigg]  \, d\tau 
    \ ,
\end{multline}
where $V(t)$ and $I(t)$ are the time histories of the imposed potential and current, respectively, and $\phi^{\rm V}$ and $\phi^{\rm C}$ are the solutions for unit voltage and current, respectively. Nonlinear gating kinetics therefore enter only through the functional dependence $I(V(t))$ but the electrochemical response itself remains additive.

\subsection{Faradaic electrodes}
\label{sec:Faradaic_main_manuscript}
We now briefly describe the dynamics of $V^{\rm M}$ when Faradaic electrodes are used instead of the blocking electrodes considered in previous sections. Unlike blocking electrodes, which prevent ionic flux, idealized Faradaic electrodes enforce ${\rho = 0}$ at the boundary by fully removing charge through reaction currents. As a result, diffuse charge layers do not form at the electrodes. Detailed calculations are presented in the SM Sec.~VI.

As with blocking electrodes, the near-field region ${r < L}$ remains unaffected by the presence of electrodes and evolves as in the unbounded case. At early times ${t < \tau_{\rm D}}$, the point-charge approximation remains valid. Furthermore, since the diffuse charge dynamics and electrode kinetics are irrelevant on early timescales, the screened response for ${r>L}$ remains well described by Eq.~\eqref{eq:point_chg_soln}.

Beyond the electrostatic regime ${t>\tau_{\rm D}}$, the dynamics are again captured by boundary layer theory. The key difference from the blocking case is that boundary layers only form at the membrane surfaces. At the Faradaic electrode surface ${z = L}$, the bulk potential directly satisfies ${\phi^{\rm b} = 0}$, replacing the boundary condition in Eq.~\eqref{eq:phi_bulk_BC_elect}. This change introduces a single characteristic speed ${v = g / (2 C_{\rm M}^{\rm eq})}$ marking the transition in dynamics at ${t = r/v}$. For intermediate times ${\tau_{\rm D} < t < r/v}$, the system exhibits a screened capacitive response
\begin{equation}
    V^{\rm M} \eq \frac{2I}{ g r}\frac{\chi}{\chi+1}\frac{t}{\tau_{\rm C}}K_0\left(\frac{\pi r}{L}\right)I_2\left(2\sqrt{\frac{t}{\tau_{\rm C}}\frac{\pi r}{L}}\right)
    e^{-\frac{t}{\tau_{\rm C}}}
    \ . \label{eq:early_time_faradaic}
\end{equation}
This expression closely resembles Eq.~\eqref{eq:early_time} for blocking electrodes; however, for Faradaic electrodes, the characteristic timescale becomes ${\tau_{\rm C} = \lambda_{\rm D} L / (D(1+\chi))}$, corresponding to $n_{\rm D}=2$ in Eq.~\eqref{eq:macro_timescale}.

For ${t > r/v}$, a qualitatively distinct regime emerges: the system relaxes to a steady state throughout the system, in contrast to the long-time diffusive behavior seen with blocking electrodes. For ${r<L}$, the steady state is given by the monopolar solution in the unbounded case. For ${r>L}$, steady-state membrane potential is given by
\begin{equation}
    V^{\rm M}(r,t) = \frac{2I}{\pi g L}\frac{\chi}{\chi+1}K_0\left(\frac{\pi r}{2 L}\right) \ , \label{eq:steady_faradaic}
\end{equation}
which decays exponentially with $r$, now with a characteristic length scale $2L/\pi$, twice that of the electrostatic regime.  This steady state arises because the transmembrane current is fully balanced by Faradaic reaction currents at the electrodes, preventing continued charge accumulation.  Thus, at any fixed position ${r > L}$, the system with Faradaic electrodes evolves through three distinct temporal regimes: electrostatic (${t < \tau_{\rm D}}$), capacitive (${\tau_{\rm D}< t < r/v}$), and steady (${t > r/v}$).

Finally, we note that the electrochemical response to simultaneous external voltage and transmembrane current remains linear and superposable, as in the blocking case. 

\subsection{Coarse-grained equivalent circuit interpretation and external current}
\label{sec:circuit}

\begin{figure}[t]
\centering
\includegraphics[width=0.87\linewidth]{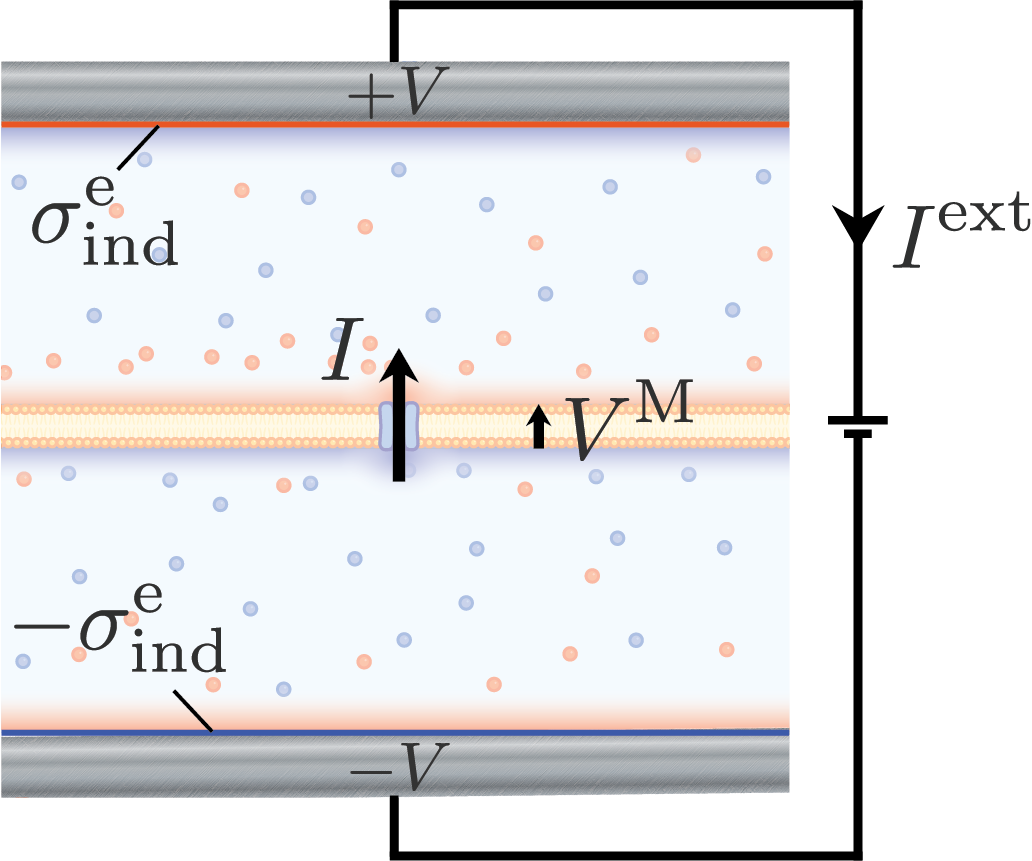}  
\caption{\textbf{External current and equivalent circuit.}  A localized membrane current \(I\) displaces equal and opposite charge onto the facing surfaces of the two blocking electrodes. As blocking electrodes do not allow diffuse charge fluxes, the charge on them can arrive only through the wire that connects the pair to the ideal battery, producing the external current \(I^{\mathrm{ext}}\) that appears in the coarse-grained balance equations.}
\label{fig_external_current}
\end{figure}

Whenever ions cross the membrane, the two metallic electrodes must develop surface charges to keep their potential difference fixed at \(2V\). This compensating charge can arise either from Faradaic currents drawn from the surrounding electrolyte, or from a time-dependent external current \(I^{\mathrm{ext}}(t)\) supplied by a potentiostat (Fig.~\ref{fig_external_current}). Let \(\sigma_{\mathrm{ind}}^{\mathrm e}(r,t)\) be the induced surface-charge
density at \({z=L}\).  A global charge balance on the upper electrode gives  
\begin{equation}
    \int_{\mathcal{S^{\rm e}}} \frac{\partial \sigma_{\rm ind}^{\rm e}}{\partial t}
     d a
    \eq
    - I^{\rm ext} +
    \int_{\mathcal{S^{\rm e}}} \te{e}_{z} \cdot \mathbf{j}_{\rho} \; d a
    \ ,
    \label{eq:induced_surf_charge_balance}
\end{equation}
where $\mathcal{S}^{\rm e}$ is the electrode surface (i.e., ${z=L}$).
Using Gauss’s law and the boundary layer solution
\begin{equation}
    \sigma^{\rm e}_{\rm ind}(r,t) 
    \eq
    \epsilon \left.\frac{\partial \phi^{\rm e}}{\partial z}\right|_{z=L} 
    \eq
    - \sigma^{\rm e}(r,t) \ ,
    \label{eq:induced_surf_charge}
\end{equation}
which indicates that the induced charge is equal and opposite to the charge stored in the electrode‐side diffuse layer.
Together with Eq.~\eqref{eq:induced_surf_charge_balance}, this connects the external current to the charging of the diffuse charge layer at the electrode.

To complete a circuit description, we next consider the global charge balance over the upper electrolyte region $\Omega$:
\begin{align*}
\int_{\Omega} \frac{\partial \rho}{\partial t} \,dv
&\eq
\int_{\mathcal{S^{\rm e}}} \frac{\partial \sigma^{\rm e}}{\partial t}\,da
+ \int_{\mathcal{S^{\rm m}}} \frac{\partial \sigma^{\rm m}}{\partial t}\,da
\\*
&\eq
I 
- \int_{\mathcal{S^{\rm e}}} \te{e}_{z} \cdot \mathbf{j}_{\rho} \; d a
\ , \numberthis
\label{eq:global_charge_bal_circuit}
\end{align*}
where $\mathcal{S}^{\rm m}$ is the membrane surface at ${z = 0}$. The first equality reflects that all charge resides only in the interfacial layers, and the second equality expresses that the net rate of charge accumulation in the electrolyte equals the charge flow due to the transmembrane current and the Faradaic current at the electrode. Combining Eqs.~\eqref{eq:memb_pot_surf_chg} and~\eqref{eq:induced_surf_charge_balance}-\eqref{eq:global_charge_bal_circuit}, we obtain 
the coarse-grained circuit equation
\begin{equation}
        I
        \eq
        I^{\mathrm{ext}}(t)+C_{\mathrm M}A^{\rm m}
        \frac{d\mathcal V^{\mathrm M}}{dt}
        \ , 
        \label{eq:global_capacitor}
\end{equation}
where ${A^{\rm m}}$ is the area of the membrane and ${\mathcal V^{\mathrm M}(t)=(A^{\rm m})^{-1}\! \int_{\mathcal{S}^{\rm m}} V^{\mathrm M}(r,t)\,da}$
is the area-averaged transmembrane potential.  Equation~\eqref{eq:global_capacitor} is the familiar circuit relation, now derived directly from the spatially resolved Poisson-Nernst-Planck framework.
Under a constant transmembrane current, charge continuously accumulates in the membrane-side diffuse layer, and $\mathcal V^{\rm M}(t)$ increases steadily until depletion effects intervene. At times ${t > \tau_{\rm C}}$, the external current approaches a steady fraction of the total ionic current: ${I^{\rm ext} = I (1 + \chi)/(2 + \chi)}$, with the remainder devoted to charging the membrane. Throughout this evolution, the transmembrane potential $V^{\rm M}(r,t)$ remains spatially nonuniform. 

Importantly, Eqs.~\eqref{eq:induced_surf_charge_balance}-\eqref{eq:global_capacitor} do not rely on any specific assumptions about electrode kinetics and are valid for both blocking and Faradaic electrodes. In the case of ideal Faradaic electrodes, however, the system reaches a steady state within ${\tau_{\rm C}}$, with ${I^{\rm ext} = I}$ and $\mathcal V^{\rm M}(t)$ to a constant value.

\section{Discussion}
\label{sec:Discussion}
\begin{figure}[hp]
\centering
\includegraphics[width=0.85\linewidth]{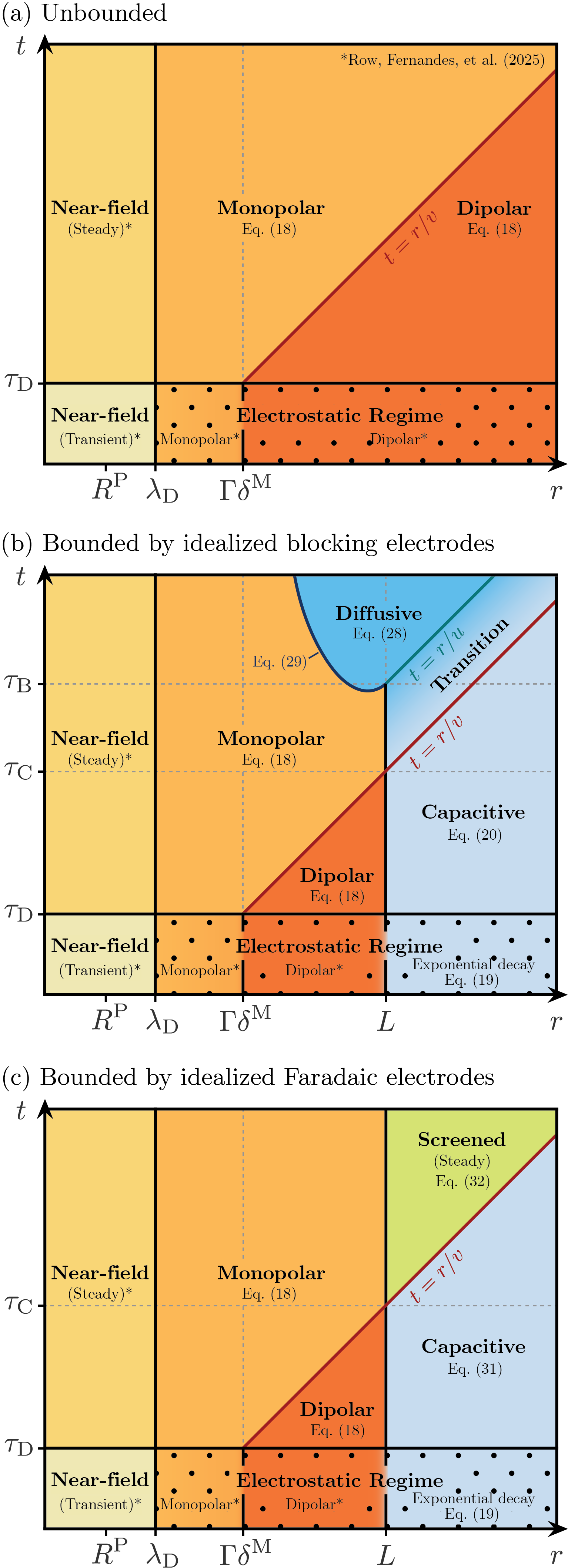}
    \caption{\textbf{Space-time diagrams}:
    The different regimes in space ($r$) and time ($t$) for the transmembrane potential $V^{\rm M}$ highlighted for (a) the case of the unbounded domain (Ref.~\cite{row2025spatiotemporal}) and the case of the electrode-bounded domain (this work) for (b) idealized blocking, and (c) idealized Faradaic electrodes.
    Axes are schematic, with scales adjusted solely to enhance visual clarity.
    }
\label{fig7_spacetime_diagram}
\end{figure}

By combining theory and simulation, we investigated the electrochemical response of a planar lipid membrane subjected simultaneously to a localized transmembrane current and a transverse step voltage applied via parallel electrodes.
We find that under physiologically relevant conditions, the effects of the imposed current and voltage are linearly superimposable. With the voltage-driven dynamics previously characterized in Ref.~\cite{farhadi2025capacitive}, we focus here on the current response, 
and uncover a hierarchy of spatiotemporal regimes that govern membrane charging in the presence of nearby electrodes. These results provide mechanistic insight into how electrochemical signals emanating from an ion channel or transporter propagate across an excitable membrane when the system is concurrently subjected to an external electric field, as in uniform field stimulation or voltage-clamp protocols.

On the nanosecond timescale (${t\lesssim\tau_{\rm D}}$), the dynamics are governed solely by electrostatics, and are well described by the point-charge approximation~\cite{row2025spatiotemporal}. At later times, the interfacial diffuse charge layers are quasistatic, while the bulk electrolyte remains electroneutral. 
For this period, we develop a boundary layer theory that captures the dynamics of the transmembrane potential $V^{\rm M}$ with high accuracy. This formulation not only enables analytical solutions and efficient numerical evaluation of $V^{\rm M}$, but also offers
conceptual clarity by revealing the physical mechanisms that drive membrane charging dynamics.

We summarize  the electrochemical response of biological membranes via ``space-time diagrams'' in Fig.~\ref{fig7_spacetime_diagram}. Regions bounded by solid lines correspond to asymptotic regimes in the radial position $r$ and time $t$. $\mbox{Panel (a)}$ recapitulates the regimes for the system in Ref.~\cite{row2025spatiotemporal}, consisting of a biological membrane in an unbounded electrolyte. Panels (b) and (c) show the response of a membrane bounded by idealized blocking and Faradaic electrodes, respectively. For distances $r$ less than the electrode separation $L$, the electrodes do not affect the dynamics.
Novel regimes emerge for ${r>L}$, beginning with electrostatic screening for ${t<\tau_{\rm D}}$ followed by the capacitive regime for ${t<r/v}$, independent of electrode kinetics. Beyond ${t=r/v}$, the electrochemical response is impacted by the nature of electrode kinetics. For blocking electrodes, a diffusive regime emerges for ${t>r/u}$, with a transitional regime separating the capacitive and diffusive responses. For Faradaic electrodes, $V^{\rm M}$ achieves a screened steady-state for ${t>r/v}$.

Charge reorganization is governed by electromigration through the bulk electrolyte, with continuous accumulation of transported charge at the membrane and electrode interfaces.
During the capacitive regime (${t < r/v}$), equal amounts of charge migrate to the membrane and electrode surfaces. If the electrodes are Faradaic, the arriving charge is consumed by reaction currents, so only the membrane accumulates charge. In contrast, for blocking electrodes, charge accumulates symmetrically at both the membrane and electrode interfaces. Regardless of electrode type, the resulting surface charges modify the electric field throughout the system. While long-ranged in the absence of electrodes, the electric field becomes radially screened when electrodes are present.

Because the membrane has a small capacitance, even modest charge deposition at its interface creates a large transmembrane potential. At later times, this potential
suppresses further charge delivery from the bulk to the membrane-side boundary layer. Consequently, beyond the capacitive regime (${t > r/v}$), charge deposition at the membrane slows, and more of the incoming charge is diverted toward the electrodes. In the Faradaic case, this redirection leads to a steady state in which all charge crossing the membrane is eventually consumed at the electrodes. For blocking electrodes, however, charge continues to accumulate in the electrode-side boundary layer, marking a transitional regime. The accumulation persists until the resulting electric field balances that induced by the membrane-side charge, eliminating axial variations in the bulk electric potential.
This axial uniformity
signals the onset of the diffusive regime, enabling 
dimensional reduction to
a radial diffusion equation and an equivalent circuit representation, as shown in Fig.~\ref{fig_BL_diffusion}.

Cable theory is a standard approach for describing electrical signal propagation along neuronal membranes \cite{cole1938,hodgkin1946,rall1962}. Originally adapted from the telegrapher’s equation to describe neuronal dynamics~\cite{Hoorweg1898}, this framework simplifies the full electrochemical system by replacing spatially distributed fields with circuit elements, effectively neglecting bulk electric fields and out-of-plane potential variations. Using the example of blocking electrodes, we demonstrate how such a circuit model can arise for a flat membrane from the underlying charge dynamics (Eqs.~\eqref{eq:diff_eq_eff}-\eqref{eq:eff_diff}). However, this reduction is not universally valid: it only emerges on timescales longer than the electrolyte charging time $\tau_{\rm B}$ (i.e., the diffusive regime), when charge accumulation at the membrane and electrode interfaces are balanced.

We further demonstrate that when considering the external circuit between the two electrodes, a coarse-grained circuit equation may be written by averaging over the electrode area, yielding Eq.~\eqref{eq:global_capacitor}. This result is consistent with standard circuit approximations of excitable membranes~\cite{hodgkin1952measurement,hille1992}. Importantly, we find that this circuit model applies only in a global sense, and does not accurately describe local membrane dynamics. In particular, for blocking electrodes, the measurable external current deviates slightly (but consistently) from the transmembrane current, due to continued charge accumulation in the membrane-side boundary layer.

Taken together, our analysis elucidates the essential physics of charge reorganization around biological membranes simultaneously subjected to localized current injection and externally applied voltage. Although centered on a minimal model system, the framework lays the groundwork for several extensions, including the incorporation of electrode kinetics, asymmetric and multivalent electrolytes, membrane surface charge, voltage-dependent transmembrane currents, and realistic curved membrane geometries~\cite{sahu2020geometry}.

\begin{acknowledgments}
J.B.F. acknowledges support from the U.S. Department of Energy, Office of Science, Office of Advanced Scientific Computing Research, Department of Energy Computational Science Graduate Fellowship under Award Number DE-SC0023112.
H.R. was partially supported by the National Science Foundation through NSF-DFG 2223407 and the Deutsche Forschungsgemeinschaft (German Research Foundation)-509322222. 
K.K.M and J.B.F are supported by Director, Office of Science, Office of Basic Energy Sciences, of the U.S. Department of Energy under contract No. DEAC02-05CH11231. 
K.S. acknowledges support from the Hellman Foundation, the McKnight Foundation, and the University of California, Berkeley. 
This research used resources of the National Energy Research Scientific Computing Center (NERSC), a U.S. Department of Energy Office of Science User Facility located at Lawrence Berkeley National Laboratory, using NERSC award BES-ERCAP0023682.
\end{acknowledgments}

\bibliography{references}

\end{document}